\documentclass{ieeeaccess}
\usepackage{cite}
\usepackage{amsmath,amssymb,amsfonts}
\usepackage{algorithmic}
\usepackage{graphicx}
\usepackage{textcomp}

\usepackage{tablefootnote}
\usepackage[para,online,flushleft]{threeparttable}
\usepackage{footnote}
\makesavenoteenv{tabular}
\makesavenoteenv{table}
\usepackage{subcaption}
\usepackage{verbatim}
\usepackage{xspace}

\usepackage{etoolbox}
\def\thevol{7}
\def\theyear{2019}
\def\myyear{2019}
\makeatletter
\patchcmd{\@evenfoot}{2016}{\myyear}{}{}
\patchcmd{\@oddfoot}{2016}{\myyear}{}{}
\makeatother

\newenvironment{ereview}{}{}

\def\BibTeX{{\rm B\kern-.05em{\sc i\kern-.025em b}\kern-.08em
    T\kern-.1667em\lower.7ex\hbox{E}\kern-.125emX}}

\begin{document}
\history{Received October 17, 2019, accepted November 11, 2019, date of publication November 21, 2019, date of current version December 5, 2019.}
\doi{10.1109/ACCESS.2019.2954957}

\title{Contextual Hybrid Session-based News Recommendation with Recurrent Neural Networks}
\author{
\uppercase{Gabriel de Souza P. Moreira}\authorrefmark{1,2},
\uppercase{Dietmar Jannach}\authorrefmark{3},
\uppercase{Adilson Marques da Cunha}\authorrefmark{1},
}

\address[1]{Department of Electrical Engineering and Computing, Instituto Tecnol\'ogico de Aeron\'autica (ITA) - S\~ao Jos\'e dos Campos, S\~ao Paulo, Brazil}
\address[2]{CI\&T - Campinas, S\~ao Paulo, Brazil}
\address[3]{Department of Applied Informatics, University of Klagenfurt, Austria}


\markboth
{Moreira \headeretal: Contextual Hybrid Session-based News Recommendation with Recurrent Neural Networks}
{Moreira \headeretal: Contextual Hybrid Session-based News Recommendation with Recurrent Neural Networks}

\corresp{Corresponding author: Gabriel de Souza P. Moreira (e-mail: gspmoreira@gmail.com).}

\begin{abstract}
Recommender systems help users deal with information overload by providing tailored item suggestions to them. The recommendation of news is often considered to be challenging, since the relevance of an article for a user can depend on a variety of factors, including the user's short-term reading interests, the reader's context, or the recency or popularity of an article. Previous work has shown that the use of Recurrent Neural Networks is promising for the next-in-session prediction task, but has certain limitations when only recorded item click sequences are used as input. In this work, we present a contextual hybrid, deep learning based approach for session-based news recommendation that is able to leverage a variety of information types. We evaluated our approach on two public datasets, using a temporal evaluation protocol that simulates the dynamics of a news portal in a realistic way. Our results confirm the benefits of considering additional types of information, including article popularity and recency, in the proposed way, resulting in significantly higher recommendation accuracy and catalog coverage than other session-based algorithms. Additional experiments show that the proposed parameterizable loss function used in our method also allows us to balance two usually conflicting quality factors, accuracy and novelty.

\end{abstract}


\titlepgskip=-15pt

\maketitle


\section{Introduction}
\label{sec:introduction}

\PARstart{R}{ecommender} Systems (RS) are nowadays widely used on modern online services, where they help users finding relevant content. Today, the application fields of recommenders range from the suggestion of items on e-commerce sites, over music recommendations on streaming platforms, to friend recommendations on social networks, where they can generate substantial business value \cite{Gomez-Uribe:2015:NRS:2869770.2843948,DBLP:conf/icis/LeeH14}.

One of the earliest application domains is the recommendation of online \textit{news} \cite{karimi2018news}. News recommendation is sometimes considered as being particularly difficult, as it has a number of distinctive characteristics \cite{Zheng:2018:DDR:3178876.3185994}. Among other challenges, news recommenders have to deal with a constant stream of news articles being published, which at the same time can become outdated very quickly. Another challenge is that the system often cannot rely on long-term user preference profiles. Typically, most users are not logged in and their short-term reading interests must be estimated from only a few logged interactions, leading to a \textit{session-based recommendation problem} \cite{QuadranaetalCSUR2018}.
Finally, like in certain other application domains, a news RS has to find the right balance between recommending only items with the highest assumed relevance and the diversity and novelty of the recommendations as a whole \cite{szpektor2013relevance,vargas2014coverage,castells2015novelty,wu2016relevance,cheng2017learning}.

In recent years, we observed an increased interest in the problem of session-based recommendation, where the task is to recommend relevant items given an ongoing user session. Recurrent Neural Networks (RNN) represent a natural choice for sequence prediction tasks, as they can learn models from sequential data. \textit{GRU4Rec} \cite{hidasi2016} was one of the first neural session-based recommendation techniques, and a number of other approaches were proposed in recent years that rely on deep learning architectures, as in \cite{Liu2018stamp, Li2017narm}.

However, as shown in \cite{jannach2017recurrent,ludewig2018evaluation,LudewigMauro2019}, neural approaches that only rely on logged item interactions have certain limitations and they can, depending on the experimental setting, be outperformed by much simpler approaches based, e.g., on nearest-neighbor techniques.

One typical way of improving the quality of the recommendations in sparse-data situations is adopt a hybrid approach and consider additional information to assess the relevance of an item \cite{paradarami2017hybrid,kim2017deep,10.7717/peerj-cs.63}. Previous approaches in the context of session-based recommendation for example used content \cite{Tuan:2017:CNS:3109859.3109900} or context information \cite{Zhang:2018:DJN:3209542.3209557} for improved recommendations.
In our work, we adopt a similar approach.

Differently from existing works, however, we consider multiple types of side information in parallel and rely on a corresponding system architecture that allows us to combine different information types.
Specifically, we adopt the general conceptual model for news recommendation that we initially proposed in \cite{moreira2018chameleon}, and base our implementation on the corresponding meta-architecture for news recommender systems called \textit{CHAMELEON} \cite{moreira2018news}. This meta-architecture was designed to address specific challenges of the news domain, like the fast decay of item relevance and extreme user- and item-cold start problems.

Going far beyond the initial analyses presented in these previous papers, we investigate, in this current work, the effects of using various information sources on different quality factors for recommendations, namely accuracy, coverage, novelty, and diversity. Furthermore, we propose a novel approach that allows us to balance potential trade-offs---e.g., accuracy vs.~novelty---depending on the specific needs of a given application.

The Research Questions (RQ) of this work are as follows:
\begin{itemize}
\item \textit{RQ1} - How does our technical approach perform compared to existing approaches for session-based recommendation?
\item \textit{RQ2} - What is the effect of leveraging different types of information on the quality of the recommendations?
\item \textit{RQ3} - How can we balance competing quality factors in our neural-based recommender system?

\end{itemize}

We answer these questions through a series of experiments based on two public datasets from the news domain. One of these datasets is made publicly available in the context of this research. Our experiments will show that (a) considering a multitude of information sources is indeed helpful to improve the recommendations along all of the considered quality dimensions and (b) that the proposed balancing approach is effective. To ensure the repeatability of our research, we publicly share the code that was used in our experiments, which not only includes the code for the proposed approach and the baselines, but also the code for data pre-processing, parameter optimization, and evaluation.

The rest of this paper is organized as follows. Next, in Section \ref{sec:previous-work}, we review existing works and previous technical approaches. In Section \ref{sec:technical-approach}, we summarize the \textit{CHAMELEON} meta-architecture and present details of our proposed method. In Section \ref{sec:experimental-evaluation}, the experimental design is described and in Section~\ref{sec:results} we present and discuss our results. The paper ends with a summary and outlook on future works in Section \ref{sec:summary-and-outlook}.

\section{Background and Related Work}
\label{sec:previous-work}
In this section, we will first review challenges of news recommendation in more detail and summarize the conceptual model for news recommendation presented in \cite{moreira2018chameleon}. We will then discuss previous approaches of applying deep learning for certain recommendation tasks.
Finally, we will briefly survey existing works on different quality factors for recommender systems.

\subsection{News Recommender Systems}
\label{subsec:related-news-recommender}
The problem of filtering and recommending news items has been investigated for more than 20 years now, see \cite{konstan1997} for an early work in this area. Technically, a variety of approaches have been put forward in these years, from collaborative filtering approaches \cite{das2007, diez2016}, to content-based methods \cite{li2011scene, capelle2012semantics, ren2013concert, ilievski2013personalized, mohallick2017exploring,bogers2007comparing}, or hybrid systems \cite{li2011scene, chu2009personalized, liu2010personalized,  rao2013personalized, lin2014personalized, li2014modeling, trevisiol2014cold, epure2017recommending}, see also \cite{karimi2018news} and \cite{ozgobek2014survey}  for recent surveys.

\subsubsection{Challenges of News Recommendation}

The main goal of personalized news recommendation is to help readers finding interesting stories that maximally match their reading interests \cite{lin2014personalized}. The news domain has, however, a number of characteristics that makes the recommendation task particularly difficult, among them the following \cite{karimi2018news,ozgobek2014survey}:

\begin{itemize}
    \item \textit{Extreme user cold-start} - On many news sites, the users are anonymous or not logged in. News portals have often very little or no information about an individual user's past behavior \cite{li2011scene, diez2016, lin2014personalized};
    \item \textit{Accelerated decay of item relevance} - The relevance of an article can decrease very quickly after publication and can also be immediately outdated when new information about an ongoing development is available. Considering the recency of items is therefore very important to achieve high recommendation quality, as each item is expected to have a short shelf life \cite{das2007, ozgobek2014survey};
    \item \textit{Fast growing number of items} - Hundreds of new stories are added daily in news portals \cite{spangher2015}. This intensifies the item cold-start problem. However, fresh items have to be considered for recommendation, even if not too many interactions are recorded for them \cite{diez2016}. Scalability problems may arise as well, in particular for news aggregators, due to the high volume of new articles being published \cite{karimi2018news, mohallick2017exploring, ozgobek2014survey};
    \item \textit{Users preferences shift} - The preferences of individual users are often not as stable as in other domains like entertainment \cite{diez2016}. Moreover, short-term interests of users can also be highly determined by their contextual situation \cite{diez2016, campos2014time, kille2013plista, ma2016user} or by exceptional situations like breaking news  \cite{epure2017recommending}.
\end{itemize}

The technical approach chosen in our work takes many of these challenges into account. In particular, it supports the consideration of short-term interests through the utilization of a neural session-based recommendation technique based on RNNs. Furthermore, factors like article recency \cite{karimi2018news, yeung2010proactive, bielikova2012effective} and general popularity \cite{10.7717/peerj-cs.63} are taken into account along with the users' context. Finally, our next-article prediction approach supports online learning in a streaming scenario \cite{jugovac2018streamingrec}, and is able, due to its hybrid nature, to recommend items that were not seen in training data.

\subsubsection{Factors Influencing the Relevance of News Items}
\begin{figure*}[h!t]
	\centering\includegraphics[width=0.75\linewidth]{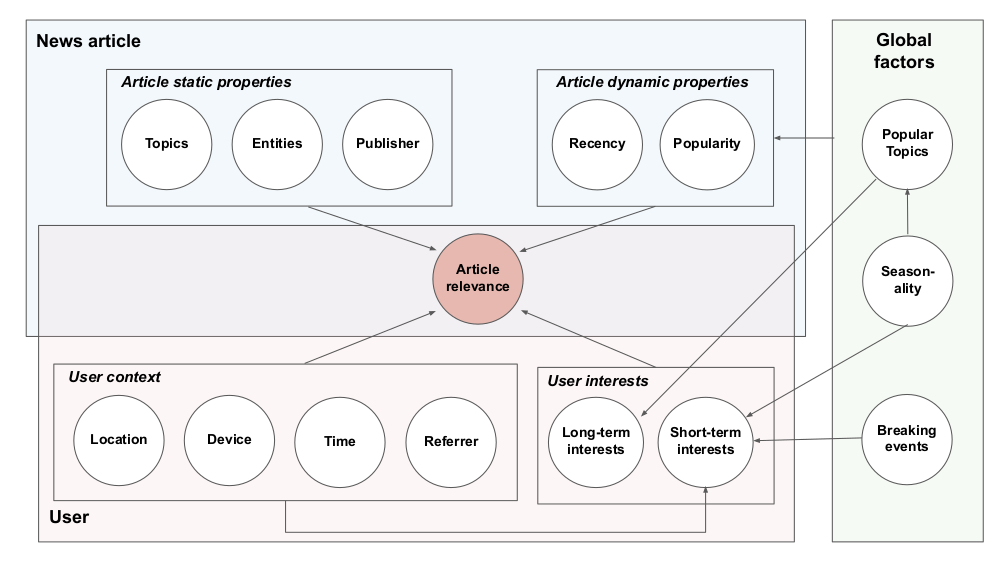}
	\caption{Conceptual model of news relevance factors}
	\label{figure:relevance-conceptual-model}
\end{figure*}

Fig.~\ref{figure:relevance-conceptual-model} shows the conceptual background of our proposed solution. In this model, a number of factors can influence the relevance of a news article for an individual user, including article-related ones, user-related ones, and what we call global factors.

With respect to \textit{article-related} factors, we distinguish between static and dynamic properties. Static properties refer to the article's content (text), its title, topic, mentioned entities (e.g., places and people) or other metadata \cite{li2011scene, said2013month}.
The reputation of the publisher can also add trust to an article \cite{lommatzsch2014real, gulla2016intricacies}.
Some news-related aspects can also dynamically change, in particular its popularity  \cite{chu2009personalized, lommatzsch2017incorporating} and recency \cite{trevisiol2014cold, gulla2016intricacies}. On landing pages of news portals, those two properties are typically the most important ranking criteria and in comparative evaluations, recommending recently popular items often shows to be a comparably well-performing strategy \cite{karimi2018news}.

When considering \textit{user-related} factors, we distinguish between the users' (short-term and long-term) interests and contextual factors. Regarding the context, their location  \cite{fortuna2010real, montes2013towards, tavakolifard2013tailored}, their device \cite{lee2007moners}, and the current time \cite{mohallick2017exploring, montes2013towards} can influence the users' short term interests, and thus the relevance of a news article \cite{mohallick2017exploring, said2013month}. In addition, the \textit{referrer} URL can contain helpful information about a user's navigation and reading context \cite{trevisiol2014cold}.

Considering the user's long-term interests can also be helpful, as some user preferences might be stable over extended periods of time \cite{diez2016}. Such interests may be specific personal preferences (e.g., chess playing) or influenced by popular global topics (e.g., on technology). In this work, we address only short-term user preferences, since we focus on scenarios where most users are anonymous. In general, however, as shown in \cite{quadrana2017personalizing}, it is possible to merge long-term and short-term interests by combining different RNNs when modeling user preferences.

Finally, there are global factors that can affect the general popularity of an item, and thus, its relevance for a larger user community. Such global factors include, for example, breaking news regarding natural disasters or celebrity news. Some topics are generally popular for many users (e.g., sports events like Olympic Games); and some follow some seasonality (e.g., political elections), which also influences the relevance of individual articles at a given point in time \cite{chu2009personalized, gulla2016intricacies, lommatzsch2017incorporating}.

\subsection{Deep Learning for Recommender Systems}
Within the last few years, deep learning methods have begun to dominate the landscape of algorithmic research in RS, see \cite{DBLP:journals/corr/ZhangYS17aa} for a recent overview. In this specific instantiation of the \textit{CHAMELEON} meta-architecture \cite{moreira2018news}, we implement two major tasks using deep learning  techniques: (a) learning article representations and (b) computing session-based recommendations.

\subsubsection{Deep Feature Extraction from  Textual Data for Recommendation}
Traditional recommendation approaches to leverage textual either use bag-of-words or TF-IDF encodings  to represent item content or meta-data descriptions \cite{melville2002content,agarwal2009,salton1988term} or they rely on topic modeling \cite{wang2011collaborative, gopalan2014content}. A potential drawback of these approaches is that they do not take word orders and the surrounding words of a keyword into account \cite{kim2016convolutional}.

Newer approaches therefore aim to extract more useful features directly from the text and use them for recommendation. Today's techniques in particular include words embeddings, paragraph vectors, Convolutional Neural Networks (CNNs), and RNNs \cite{bansal2016ask}. Kim \textit{et al.}~\cite{kim2016convolutional}, for example, proposed Convolutional Matrix Factorization (\textit{ConvMF}), which combines a CNN with Probabilistic Matrix Factorization to leverage information from user reviews for rating prediction.

Similarly, Seo \textit{et al.}~\cite{seo2017interpretable} aim to jointly model user preferences and item properties using a CNN, using a local and global attention mechanism.

Using a quite different approach, Bansal \textit{et al.}~\cite{bansal2016ask} used an RNN to learn representations from the textual content of scientific papers. Besides predicting ratings for a given article, they used multi-task learning to predict also item metadata such as genres or item tags from text.

Our work shares similarities with these previous works in that we extract features using deep learning, in our case with a CNN, based on pre-trained word embeddings. However, instead of predicting ratings, our approach learns a representation of an article's content by training a separate neural network for a side task---predicting article metadata attributes based on its text.

Differently from \cite{bansal2016ask}, we also do not rely on an end-to-end model to extract features and to recommend items. Instead, we rely on two different modules in order to ensure scalability, given the often huge amount of recorded user interactions and news articles published every day \cite{ozgobek2014survey, karimi2018news}. The details of our approach will be discussed in Section \ref{sec:technical-approach}.

\subsubsection{Deep Learning for Session-based Recommendation}
RNNs are a natural choice for session-based recommendation scenarios as they are able to model sequences in datasets \cite{donkers2017sequential}. \textit{GRU4Rec}, proposed by Hidasi \textit{et al.}~\cite{hidasi2016}, represents one of the earliest approaches in that context. In their approach, the authors specifically use Gated Recurrent Units (GRU) to be better able to deal with longer sessions and the vanishing gradient problem of RNNs. Later on, a number of improvements were published by the authors in terms of more effective loss functions  \cite{hidasi2018recurrent}.

One limitation of \textit{GRU4Rec} in the news domain is that the method can only recommend items that appeared in the training set, because it is trained to predict scores for a fixed number of items. Another potential limitation is that RNN-based approaches that only use item IDs for learning with no side information might not be much better or even worse in terms of prediction accuracy than simpler approaches. Detailed analyses of this phenomenon can be found in \cite{jannach2017recurrent,jugovac2018streamingrec,ludewig2018evaluation}.

A number of works, however, exist that propose RNN-based approaches that use additional side information about the user's context or the items. In \cite{hidasi2016parallel}, for example, the authors extended \textit{GRU4Rec} to additionally use image and textual descriptions of the items. Like in our work, they did not apply an integrated end-to-end approach, but extracted image features independently by using transfer learning from a pre-trained network \cite{szegedy2015going} and used simple \textit{TF-IDF} vectors for textual representations.

Contextual information was used in combination with RNNs, for example, in \cite{smirnova2017contextual} or \cite{twardowski2016modelling}. In \cite{smirnova2017contextual}, the authors consider not only the sequence of events when making predictions but also the type of the event, the time gaps between events, or the time of the day of an event, leading to what they call Contextual Recurrent Neural Networks for Recommendation (\textit{CRNN}). Similarly, Twardowski \cite{twardowski2016modelling} considers time as a contextual factor that is combined with item information within a hybrid approach.

A work that has certain similarities with ours in terms of the recommendation approach is the Recurrent Attention \textit{DSSM} (\textit{RA-DSSM}) model by Kumar \cite{kumar2017}.

The \textit{RA-DSSM} is an adaptation for the news domain of the \textit{ Multi-View Deep Neural Network (MV-DNN)}, which extended the Deep Structured Semantic Model (\textit{DSSM}) \cite{huang2013learning} information retrieval architecture to recommender systems.  The \textit{(MV-DNN)} maps users and items to a shared semantic space and recommend items that have the highest similarity with the users in the mapped space.

Technically, the authors use a bidirectional \textit{LSTM} layer with an attention mechanism \cite{bahdanau2014neural}. Similarly to our instantiation of the \textit{CHAMELEON} framework, they rely on RNNs as a base building block, use embeddings to represent textual content and implement a similarity-based loss function derived from \textit{MV-DNN}. The \textit{CHAMELEON} meta-architecture however, as will be discussed in Section~\ref{subsec:chameleon-meta-architecture}, lives at a higher level of abstraction than the specific \textit{RA-DSSM} model.

Our solution also differs from \textit{RA-DSSM} in a number of other dimensions. \textit{RA-DSSM} for example uses \textit{doc2vec} embeddings \cite{le2014distributed} to represent content, while we propose a specific neural architecture to learn textual representations based on pre-trained word embeddings for improved accuracy.

Furthermore, the \textit{RA-DSSM} does not use any contextual information about users or articles, which may limit its accuracy in cold-start scenarios that are common in news recommendation. Article recency and popularity were not considered in their model as well. Additionally, we use a temporal evaluation protocol to emulate a more realistic scenario, described in Section~\ref{sec:eval_methodology}, while their experiments do not mimic the dynamics of a news portal.

\subsubsection{Deep Reinforcement Learning for News Recommendation}
\begin{ereview}
Reinforcement learning is an alternative technical approach for recommending online news, and often multi-arm (contextual) bandit models were applied for the task \cite{DBLP:conf/www/LiCLS10}.
In \cite{Zheng:2018:DDR:3178876.3185994}, the authors propose a novel deep reinforcement learning technique for news recommendation. Differently from our problem setting, the authors focus on \emph{session-aware} recommendations, where longer-term information about individual users is available. Similarly to our work, however, the approach proposed in \cite{Zheng:2018:DDR:3178876.3185994} relies on a number of features that we also used in our models, e.g., article metadata, recent click counts, and context features. In their problem setting with longer-term models, the authors in addition included a number of user-related pieces of information, which are typically not available in session-based recommendation task, e.g., preferences regarding different content categories over longer periods of time.
\end{ereview}

\subsection{Balancing Accuracy and Novelty in Recommender Systems}
It is known for many years that prediction accuracy is not the only factor that determines the success of a recommender. Other quality factors discussed in the literature are, e.g., novelty, catalog coverage, diversity \cite{di2017adaptive}, or reliability \cite{bobadilla2018reliability}. In the context of news recommendation, the aspect of novelty is particularly relevant to avoid a ``rich-get-richer'' phenomenon where a small set of already popular articles get further promoted through recommendations and less popular or more recent items rarely make it into a recommendation list.

The novelty of a recommended item can be defined in different ways, e.g., as the non-obviousness of the item suggestions \cite{Herlocker2004}, or in terms of how different an item is with respect to what has already been experienced by a user or the community \cite{vargas2011rank}. Recommending solely novel or unpopular items can, however, be of limited value when they do not match the users' interests well. Therefore, the goal of a recommender is often to balance these competing factors, i.e., make somewhat more novel and thus risky recommendations, while at the same time ensuring high accuracy.

In the literature, a number of ways have been proposed to \textit{quantify} the degree of novelty, including alternative ways of considering popularity information \cite{zhou2010solving} or the distance of a candidate item to the user's profile \cite{karimi2018news, nakatsuji2010classical, rao2013taxonomy}. In \cite{vargas2011rank}, the authors propose to measure novelty as the opposite of popularity of an item, under the assumption that less popular (long-tail) items are more likely to be unknown to users and their recommendation will, hopefully, lead to higher novelty levels. In our work, we will also consider the novelty of the recommendations and adopt existing novelty metrics from the literature.

Regarding the treatment of trade-off situations, different technical approaches are possible. One can, for example, try to re-rank an accuracy-optimized recommendation list, either to meet globally defined quality levels \cite{adomavicius2011improving} or to achieve recommendation lists that match the preferences of individual users \cite{JugovacJannachLerche2017eswa}. Another approach is to vary the weights of the different factors to find a configuration that leads to both high accuracy and good novelty \cite{garcin2013personalized}.

Finally, one can try to embed the consideration of trade-offs within the learning phase, e.g., by using a corresponding regularization term. In \cite{coba2018novelty}, the authors propose a method called Novelty-aware Matrix Factorization (\textit{NMF}), which tries to simultaneously recommend accurate and novel items. Their proposed regularization approach is pointwise, meaning that the novelty of each candidate item is considered individually.

In our recommendation approach, we consider trade-offs in the regularization term as well. Differently, from \cite{coba2018novelty}, however, our approach is not focused on matrix factorization, but rather on neural models that are derived from the \textit{DSSM}. Furthermore, the objective function in our work uses a listwise ranking approach to learn how to enhance the novelty level of the top-n recommendations.

\section{Technical Approach}
\label{sec:technical-approach}
The work presented in this paper is based on an instantiation of the \textit{CHAMELEON} meta-architecture, which we presented in an initial version in \cite{moreira2018news}. The meta-architecture is designed for building session-based news recommendation systems, which are context-aware and can leverage additional content information.

We will discuss this meta-architecture next in Section~\ref{subsec:chameleon-meta-architecture}. Afterwards, in Section~\ref{subsec:chameleon-instantiation}, we provide information about the specific instantiation used for our experiments. Finally, in Section~\ref{subsec:loss-function} propose a novel technical approach to balance accuracy and novelty based on a parameterizable loss function.

\subsection{The CHAMELEON Meta-Architecture}
\label{subsec:chameleon-meta-architecture}

The \textit{CHAMELEON} meta-architecture was designed to deal with some of the specific requirements of news recommendation, as outlined in Section~\ref{subsec:related-news-recommender}. Generally, when building a news recommender system, one has several design choices regarding the types of data that are used, the chosen algorithms, and the specific network architecture when relying on deep learning approaches. With \textit{CHAMELEON}, we provide an architectural abstraction (a ``meta-architecture''), which contains a number of general building blocks for news recommenders and which can be instantiated in various ways, depending on the particularities of the given problem setting.

\begin{figure*}[h!t]
	\centering\includegraphics[width=0.75\linewidth]{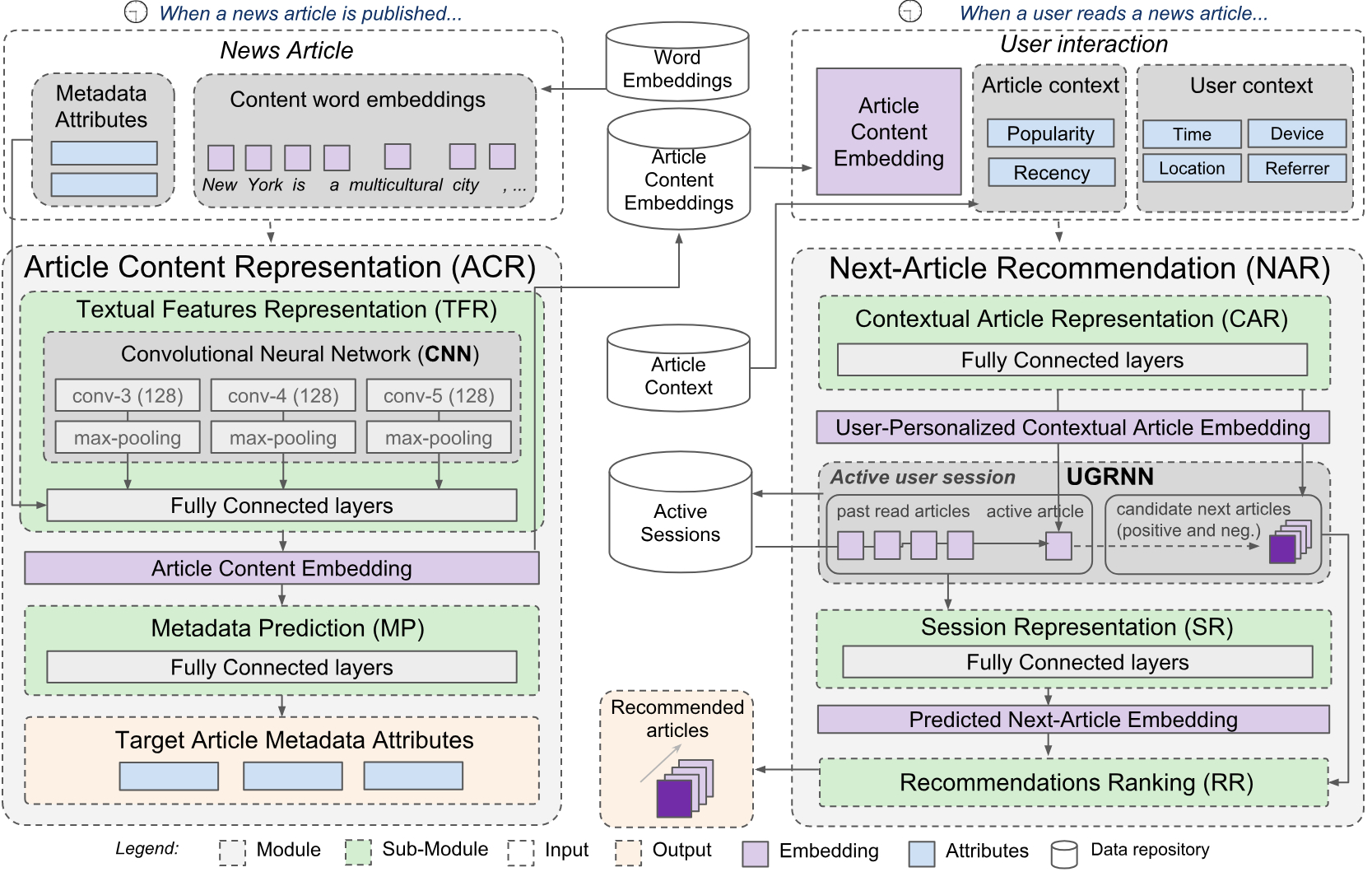}	\caption{An architecture instantiation of \textit{CHAMELEON}}
	\label{figure:chameleon_instantiation}
\end{figure*}

Fig.~\ref{figure:chameleon_instantiation} shows the \textit{main building blocks} of the meta-architecture and also sketches how it was instantiated for the purpose of this research. At its core, \textit{CHAMELEON} consists of two complementary modules, with independent life cycles for training and inference:
\begin{itemize}
  \item The \textit{Article Content Representation} (\textit{ACR}) module used to learn a distributed representation (an embedding) of the articles' content; and
  \item The \textit{Next-Article Recommendation} (\textit{NAR}) module responsible to generate next-article recommendations for ongoing user sessions.
\end{itemize}

In a \textit{CHAMELEON}-based architecture, the \textit{ACR} module learns an \textit{Article Content Embedding} for each article independently from the recorded user sessions. This is done for scalability reasons, because training user interactions and articles in a joint process would be computationally very expensive, given the typically large amount of recorded user interactions. Instead, the internal model is trained for a side classification task---predicting target metadata attributes (e.g. news category, topic, tags) of an article.

After training, the learned \textit{Article Content Embeddings} are stored in a repository for further usage by the \textit{Next Article Recommendation} module.

The \textit{NAR} module, which provides recommendations for active sessions, is designed as a hybrid recommender system, considering both the recorded user interactions and the content of the news articles. It is also context-aware in that it leverages information about the usage context, e.g., location, device, previous clicks in the session, and the article's context --- popularity and recency -- which quickly decay over time. All these inputs are combined by feed-forward layers to produce what we call a \textit{User-Personalized Contextual Article Embedding}. As a result, we obtain individualized article embeddings, whose representations depend on the user's context and other factors such as the article's current popularity and recency.

Generally, considering these additional factors can be crucial for the effectiveness of the recommendations, in particular as previous work has shown that RNNs without side information are often not much better than relatively simple algorithms \cite{jannach2017recurrent, ludewig2018evaluation}. Additional details about the \textit{CHAMELEON} meta-architecture can be found in \cite{moreira2018news}.

\subsection{Specific Instantiation}
\label{subsec:chameleon-instantiation}

For the experiments conducted in this work, we used an instantiation of the \textit{ACR} module that is similar to the one from \cite{moreira2018news}. Specifically, we extract features from textual content with a CNN. The \textit{Article Content Embeddings} were trained to predict target article metadata attributes. In order to support multiple target attributes, a new loss function was designed to compute a weighted sum of classification losses for single-label (\textit{softmax cross-entropy}) and multi-label attributes (\textit{sigmoid cross-entropy}), e.g., tags and keywords. The architecture of the \textit{ACR} module and the training protocol is described in more detail in \cite{moreira2018news}. The input and output features for each dataset used in the experiments will be presented in Section~\ref{sec:datasets}.

Furthermore, the \textit{NAR} module was instantiated with some improvements compared to \cite{moreira2018news}. Generally, the \textit{NAR} module uses RNNs to model the sequence of user interactions. We empirically tested different RNN cells, like variations of \textit{LSTM} \cite{hochreiter1997long} and \textit{GRU} \cite{chung2014empirical}, whose results were very similar. At the end, we selected the \textit{\textit{Update Gate RNN} (UGRNN)} cell \cite{collins2016capacity}, as it led to slightly higher accuracy. The \textit{UGRNN} architecture is a compromise between \textit{LSTM}/\textit{GRU} and a vanilla RNN. In the \textit{UGRNN} architecture, there is only one additional gate, which determines whether the hidden state should be updated or carried over \cite{collins2016capacity}.
Adding a new (non bi-directional) RNN
layer on top of the previous one also led to some accuracy improvement.

In a first step, the \textit{NAR} module derives what we call a \textit{User-Personalized Contextual Article Embedding} as described above. Specifically, in our instantiation, we consider the recent popularity of an article (e.g., by considering the clicks within the last hour) and its recency in terms of hours since its publication. As the user's context, we consider the time, location,  device, and referrer type in case this information is available. The overall training phase of the \textit{NAR} module then consists in learning a model that relates these \textit{User-Personalized Contextual Article Embeddings} of the recommendable articles with the \textit{Predicted Next Article Embeddings}, based on representations learned by the RNN from past session information.

Specifically, the optimization goal is to \textit{maximize the similarity} between the \textit{Predicted Next-Article Embedding} and the \textit{User-Personalized Contextual Article Embedding} corresponding to the next article actually read by the user in his or her session (positive sample), whilst minimizing its similarity with negative samples (articles not read by the user during the session)\footnote{The approach is inspired by the \textit{DSSM} \cite{huang2013learning} and by later works that applied the idea for recommender systems \cite{elkahky2015multi,kumar2017,song2016multi} and which use a ranking loss function based on the similarity of embeddings.}.
Using this strategy, a newly published article can be immediately recommended, as soon as its \textit{Article Content Embedding} is added to the repository. Details regarding the optimization problem are described next.

\subsection{A Parameterizable Loss Function to Balance Accuracy and Novelty}
\label{subsec:loss-function}

In this section, we describe the loss function of the \textit{NAR} module, designed to optimize for accuracy (Section~\ref{sec:acc_loss}) and a newly proposed extension to balance accuracy and novelty (Section~\ref{sec:new_loss_function}).

\subsubsection{Optimizing for Recommendation Accuracy}
\label{sec:acc_loss}

Formally, we can describe the method for optimizing prediction accuracy as follows. The inputs for the \textit{NAR} module, described later in Table~\ref{tab:nar-features}, are represented by ``$i$'' as the article ID, ``$uc$'' as the user context, ``$ax$'' as the article context, and ``$ac$'' as the article textual content. Based on those inputs, we define ``$cae =\Psi(i, ac, ax, uc)$'' as the \textit{User-Personalized Contextual Article Embedding}, where $ \Psi(\cdot) $ represents a sequence of fully-connected layers with non-linear activation functions to combine the inputs for the RNN.

The symbol $s$ stands for the user session (sequence of articles previously read, represented by their $ cae $ vectors), and ``$ nae=\Gamma (s) $'' denotes the \textit{Predicted Next-Article Embedding}, where $ \Gamma(\cdot) $ is the output embedding predicted by the RNN as the next article.

In \eqref{eq:relevance}, the function $R$ describes the relevance of an item $ i $ for a given user session $ s $ as the similarity between the $nae$ vector predicted as the next-article for the session and the $cae$ vectors from the recommendable articles.

\begin{equation} \label{eq:relevance}
R(s, i) = \text{sim}(nae, cae)
\end{equation}

In the \textit{NAR} module instantiation presented in \cite{moreira2018news}, the $\text{sim}(\cdot) $ function was simply the cosine similarity. For this study, it was instantiated as the element-wise product of the embeddings, followed by a number of feed-forward layers. This setting allows the network to flexibly learn an arbitrary matching function:

\begin{equation} \label{eq:sim}
\text{sim}(nae, cae) = \phi(nae \odot cae),
\end{equation}

where $ \phi(\cdot) $ represents a sequence of fully-connected layers with non-linear activation functions, and where the last layer outputs a single scalar representing the relevance of an article as the predicted next article. In our study, $ \phi(\cdot) $ consisted of a sequence of 4 feed-forward layers with a \textit{Leaky ReLU} activation function \cite{maas2013rectifier}, with 128, 64, 32, and 1 output units.

The ultimate task of the \textit{NAR} module is to produce a ranked list of items (top-n recommendation) that we assume the user will read next\footnote{This corresponds to a typical next-click prediction problem.}. Using $i \in D$ to denote the set of all items that can be recommended, we can define a ranking-based loss function for a problem setting as follows. The goal of the learning task is to maximize the similarity between the predicted next article embedding ($nae$) for the session and the $cae$ vector of the next-read article (positive sample, denoted as $ i^+ $), while minimizing the pairwise similarity between the $nae$ and the and $cae$ vectors of the negative samples $i^- \in D^-$. i.e., those that were not read by the user in this session.
Since $D$ can be large in the news domain, we approximate it through a set $D'$, which is the union of the unit set of the read articles (positive sample) $ \{i^+\} $ and a set with random negative samples from $D^-$.

As proposed in \cite{huang2013learning}, we compute the posterior probability of an article being the next one given an active user session with a \textit{softmax} function over the relevance scores:

\begin{ereview}
\begin{equation} \label{eq:prob_click}
P(i \mid s, D') = \frac{ \text{exp}(\gamma R(i, s))}{\sum_{\forall i' \in D'}{ \text{exp}(\gamma R(i', s)})}
\end{equation}
\end{ereview}

where $\gamma$ is a smoothing factor (usually referred to as \textit{temperature}) for the softmax function, which can be trained on a held-out dataset or which can be empirically set.

Using these definitions, the model parameters $\theta$ in the \textit{NAR} module are estimated to maximize the accuracy of the recommendations, i.e, the likelihood of correctly predicting the next article given a user session. The corresponding loss function to be minimized, as proposed in \cite{huang2013learning}:

\begin{ereview}
\begin{equation} \label{eq:acc_loss}
\text{accuracy\_loss}(\theta) = \frac{1}{|C|} \sum_{ (s, i^+, D') \in C} -\text{log} (P(i^+ \mid s, D')),
\end{equation}
\end{ereview}

where $ C $ is the set of user clicks available for training, whose elements are triples of the form $ (s,i^+, D')$.

Since $ accuracy\_loss(\theta) $ is differentiable w.r.t.\,to $ \theta $ (the model parameters to be learned), we can use back-propagation on gradient-based numerical optimization algorithms in the \textit{NAR} module.

\subsubsection{Balancing Recommendations Accuracy and Novelty}
\label{sec:new_loss_function}
In order to incorporate the aspect of novelty of the recommendations directly in the learning process, we propose to include a novelty regularization term in the loss function of the \textit{NAR} module. This regularization term has a hyper-parameter which can be tuned to achieve a balance between novelty and accuracy, according to the desired effect for the given application. Note that this approach is not limited to particular instantiations of the \textit{CHAMELEON} meta-architecture, but can be applied to any other neural architecture which takes the article's recent popularity as one of the inputs and uses a \textit{softmax} loss function for training \cite{huang2013learning}.

In our approach, we adopt the novelty definition proposed in \cite{vargas2011rank, vargas2015thesis}, which is based on the inverse popularity of an item. The underlying assumption  of this definition is that less popular (long-tail) items are  more likely to be unknown to users and their recommendation will lead to higher novelty levels \cite{karimi2018news}.

The proposed novelty component therefore aims to bias the recommendations of the neural network toward more novel items.
The corresponding regularization term is based on listwise ranking, optimizing the novelty of a recommendation list in a single step. The positive items (actually clicked by the user) are not penalized based on their popularity, only the negative samples. The novelty of the negative items is weighted by their probabilities to be the next item in the sequence (computed according to \eqref{eq:prob_click} in order to push those items to the top of the recommendation lists 
that are both novel and relevant.

Formally, we define the novelty loss component as:

\begin{equation} \label{eq:nov_reg}
\begin{split}
\text{nov\_loss}(\theta) = & \frac{1}{|C|} \sum_{ (s, i^+, D'^{-}) \in C}  \\&  \frac{\sum_{i \in D'^{-}} P(i \mid s, D'^{-}) * \text{novelty}(i)}
{\sum_{i \in D'^{-}} P(i \mid s, D'^{-})},
\end{split}
\end{equation}

where $C$ is the set of recorded click events for training, $ D'^{-} $ is a random sample of the negative samples, not including the positive sample as in the accuracy loss function \eqref{eq:acc_loss}. The novelty values of the items are weighted by their predicted relevance $ P(i \mid s, D'^{-}) $ in order to push both novel and relevant items towards the top of the recommendations list.

The novelty metric in \eqref{eq:novelty} is defined based on the \textit{recent normalized popularity} of the items. The negative logarithm in \eqref{eq:novelty} increases the value of the novelty metric for long-tail items. The computation of the \textit{normalized popularity} sums up to 1.0 for all recommendable items (set $I$), as shown in \eqref{eq:rec-norm-pop}. Since we are interested in the recent popularity, we only consider the clicks an article has received within a time frame (e.g., in the last hour), as returned by the function $\text{recent\_clicks}(\cdot)$:

\begin{equation} \label{eq:novelty}
\text{novelty}(\text{i}) = -\text{log}_2(\text{rec\_norm\_pop}(i) + 1),
\end{equation}

\begin{equation} \label{eq:rec-norm-pop}
\text{rec\_norm\_pop}(i) = \frac{\text{recent\_clicks}(i)}{\sum_{j \in I} \text{recent\_clicks}(j)}
\end{equation}

\paragraph{Complete Loss Function}

The complete loss function proposed in this work combines the objectives of accuracy and novelty:

\begin{equation}
\label{eq:final_loss}
L(\theta) = \text{accuracy\_loss}(\theta) - \beta * \text{nov\_loss}(\theta) ,
\end{equation}

where $\beta $ is the tunable hyper-parameter for novelty. Note that the novelty loss term is subtracted from the accuracy loss, as this term is higher when more novel items are recommended. The values for $\beta $ can either be set based on domain expertise or be tuned to achieve the desired effects.

\section{Experimental Evaluation}
\label{sec:experimental-evaluation}

We conducted a series of experiments to answer the research questions described above. In the context of \textit{RQ1}, our goal was to compare our method (\textit{CHAMELEON}) with existing session-based recommenders in the news domain. For \textit{RQ2}, we try to understand the effects of leveraging different types of information on the quality of the recommendations. Finally, \textit{RQ3} addresses the effectiveness of our approach on balancing the accuracy and novelty trade-off.

In this section, we first discuss our experimental design, including the used datasets and the evaluation approach. The results of the evaluation will be discussed later in Section  \ref{sec:results}.

\subsection{Datasets}
\label{sec:datasets}
We use two public news portals datasets for our evaluation. The datasets contain recorded user interactions and information about the published articles:

\begin{itemize}
    \item \textit{Globo.com} (\textit{G1}) dataset -  Globo.com is the most popular media company in Brazil. This dataset was originally shared by us in \cite{moreira2018news}. With this work, we publish a second version\footnote{https://www.kaggle.com/gspmoreira/news-portal-user-interactions-by-globocom}, which also includes contextual information. The dataset was collected from the G1 news portal, which has more than 80 million unique users and publishes over 100,000 new articles per month;
    \item \textit{SmartMedia Adressa dataset} - This dataset contains approximately 20 million page visits from a Norwegian news portal \cite{gulla2017adressa}. In our experiments we used the full dataset, which is available upon request\footnote{http://reclab.idi.ntnu.no/dataset}, and includes article text and click events of about 2 million users and 13,000 articles.
\end{itemize}

Both datasets include the textual content of the news articles, article metadata (such as publishing date, category, and author), and logged user interactions (page views) with contextual information. Since we are focusing on session-based news recommendations and short-term users preferences, it is not necessary to train algorithms for long periods. Therefore, and because articles become outdated very quickly, we have selected for the experiments all available user sessions from the first 16 days for both datasets.

In a pre-processing step, like in \cite{ludewig2018evaluation, epure2017recommending, twardowski2016modelling}, we organized the data into sessions using a 30 minute threshold of inactivity as an indicator of a new session.
Sessions were then sorted by timestamp of their first click. From each session, we removed repeated clicks on the same article, as we are not focusing on the capability of algorithms to act as reminders as in \cite{LercheJannachEtAl2016}. Sessions with only one interaction are not suitable for next-click prediction and were discarded. Sessions with more than 20 interactions (stemming from \textit{outlier} users with an unusual behavior or from bots) were truncated.

The characteristics of the resulting pre-processed datasets are shown in Table~\ref{tab:datasets}. Coincidentally, the datasets are similar in many statistics, except for the number of articles. For the \textit{G1} dataset, the number of recommendable articles (clicked by at least one user) is much higher than for the Adressa dataset. The higher Gini index of the articles' popularity distribution also indicates that the clicks in the Adressa dataset are more biased to popular articles, leading to a higher inequality in clicks distribution than for the G1 dataset.

\begin{table}[h!t]
\centering
\caption{Statistics of the datasets used for the experiments.}
\label{tab:datasets}
\vspace{10pt}
\begin{tabular}{p{4cm}rr}
\hline
 & \textit{Globo.com (G1)}
 & \textit{Adressa} \\ \hline
Language  & Portuguese & Norwegian  \\ 
Period (days)  & 16 & 16 \\  
\# users   & 322,897 & 314,661  \\ 
\# sessions & 1,048,594 & 982,210 \\  
\# clicks  & 2,988,181 & 2,648,999 \\  
\# articles   & 46,033 & 13,820  \\ 

Avg. sessions length (\# clicks / \# sessions)  & 2.84  & 2.70  \\
Gini index (of the article pop. distribution) &  0.952 &  0.969  \\
\hline
\end{tabular}
\end{table}

\subsection{Compared Recommendation Approaches}
This section describes the implementation of a specific instantiation of \textit{CHAMELEON} and of a number of baseline techniques.

\subsubsection{CHAMELEON---Implementation Specifics}
This instantiation of the \textit{CHAMELEON} meta-architecture, presented in Fig.~\ref{figure:chameleon_instantiation}, was implemented using TensorFlow \cite{abadi2016tensorflow}, a popular Deep Learning framework. We publish the source code for our neural architecture and for the baseline methods to make our experiments reproducible\footnote{https://github.com/gabrielspmoreira/chameleon\_recsys}.

The \textit{Article Content Embeddings} were trained by the \textit{ACR} module, whose input and target features for the classifier are described in Table~\ref{tab:acr-features}. Within the \textit{Next Article Recommendation (NAR)} module, rich features were extracted from the user interactions logs, as detailed in Table~\ref{tab:nar-features}. The features were prepared to be used as input for both the \textit{ACR} and \textit{NAR} modules as follows.

Categorical features with low cardinality (i.e., with less than 10 distinct values) were one-hot encoded and features with high cardinality were represented as trainable embeddings. Numerical features were standardized with z-normalization. The dynamic features \textit{Novelty} and \textit{Recency} were normalized based on a sliding window of the recent clicks (within the last hour), so that they can accommodate both repeating changes in their distributions over time, e.g., within different periods of the day, and abrupt changes in global interest, e.g., due to breaking news.

\begin{table}[!htbp]
\begin{threeparttable}
\centering
\caption{Features used by the \textit{Article Content Representation (ACR)} module.}

\label{tab:acr-features}
\vspace{10pt}
\footnotesize
\begin{tabular}{p{2cm}lp{4.5cm}}
\hline
 \textit{Features} &
 \textit{Type} &
 \textit{Description} \\
\hline
\multicolumn{3}{p{6cm}}{\textbf{Input features}}\\ \hline
Textual Content & Emb. & Article text represented as a sequence of word embeddings, pre-trained for the language of the dataset\tnote{1}. \\
Concepts, Entities, Locations, Persons & Categ. & Lists of categorical values extracted with NLP-techniques by Adressa \cite{gulla2017adressa}. Available only for the Adressa dataset. \\
\hline
\multicolumn{3}{p{6cm}}{\textbf{Target features}}\\ \hline
Category & Categ. & The category of the article, defined by editors. \\
Keywords* & Categ. & Human-labeled keywords for the Adressa dataset. \\
\hline
\end{tabular}

\begin{tablenotes}\footnotesize
\item[1] Portuguese: Pre-trained Word2Vec \textit{skip-gram} model (300 dimensions) available at http://nilc.icmc.usp.br/embeddings; Norwegian: a \textit{skip-gram} model (100 dimensions) available at http://vectors.nlpl.eu/repository (model \#100).
\end{tablenotes}

\end{threeparttable}
\end{table}

\begin{table*}[!htbp]
\begin{threeparttable}
\centering
\caption{Features used by the \textit{Next-Article Recommendation (NAR) } module}
\label{tab:nar-features}
\vspace{10pt}
\footnotesize
\begin{tabular}{p{2.5cm}llp{10cm}}
\hline
 \textit{Group} &
 \textit{Features} &
 \textit{Type} &
 \textit{Description} \\
\hline
\multicolumn{4}{p{8cm}}{\textbf{Dynamic article features}}\\ \hline
Article Context & Novelty & Num. & The novelty of an article, computed based on its normalized recent popularity, as described in \eqref{eq:novelty}.  \\
 & Recency & Num. & Computed as the logarithm of the elapsed days (with hours represented as the decimal part) since an article was published: \begin{ereview} $ \text{log}_2 $((current\_date - published\_date)+1). \end{ereview} \\ \hline
\multicolumn{4}{p{8cm}}{\textbf{Static article features}}\\ \hline
Id & Id & Emb. & Trainable embeddings for article IDs. \\
Content & ACE & Emb. & The \textit{Article Content Embedding} representation learned by the ACR module. \\
Metadata & Category & Cat. & Article category \\
 & Author * & Cat. & Article author \\ \hline
\multicolumn{4}{p{8cm}}{\textbf{User context features}}\\ \hline
Location & Country, Region, City* & Categ. & Estimated location of the user \\
Device & Device type & Categ. & Desktop, Mobile, Tablet, TV** \\
 & OS & Categ. & Device operating system\\
 & Platform** & Categ. & Web, mobile app\\
Time & Hour of the day & Num. & Hour encoded as cyclic continuous feature (using sine and cosine)\\
 & Day of the week & Num. & Day of the week \\
Referrer & Referrer type & Categ. & Type of referrer: e.g., direct access, internal traffic, search engines, social platforms, news aggregators\\
\hline
\end{tabular}

\begin{tablenotes}\footnotesize
\item [*] Only available for the Adressa dataset.
\item [**] Only available for the G1 dataset.
\end{tablenotes}

\end{threeparttable}
\end{table*}

\subsubsection{Baseline Methods}
In our experiments, we consider (a) different variants of our instantiation of the \textit{CHAMELEON} meta-architecture to assess the value of considering additional types of information and (b) a number of session-based recommender algorithms, described in Table~\ref{tab:baselines}. While \begin{ereview}some of\end{ereview} the chosen baselines appear conceptually simple, recent work has shown that some of them are able to outperform very recent neural approaches for session-based recommendation tasks \cite{jannach2017recurrent,ludewig2018evaluation,jugovac2018streamingrec}. Furthermore, the \begin{ereview} simple
\end{ereview}methods, unlike  \begin{ereview}neural-based\end{ereview} approaches, can be continuously updated over time and take newly published articles into account.

\begin{table}[!htbp]
\begin{threeparttable}
\centering
\caption{Baseline session-based recommender algorithms used in the experiments.}
\label{tab:baselines}
\vspace{10pt}
\footnotesize
\begin{tabular}{p{2.5cm}p{5cm}}
\hline
\multicolumn{2}{p{8cm}}{Neural Methods}\\ \hline

\begin{ereview}\textit{GRU4Rec}\end{ereview} & \begin{ereview} A landmark neural architecture using RNNs for session-based recommendation \cite{hidasi2016}. For this experiment, we used the \textit{GRU4Rec} v2 implementation, which includes the improvements reported in \cite{hidasi2018recurrent}.\tnote{1} We furthermore improved the algorithm's negative sampling strategy for the scenario of news recommendation.\tnote{2}\end{ereview} \\ \hline

\begin{ereview}\textit{SR-GNN} \end{ereview}& \begin{ereview} A recently published state-of-the-art architecture for session-based recommendation based on Graph Neural Networks. In \cite{wu2019session}, the authors reported superior performance over other neural architectures such as \textit{GRU4Rec} \cite{hidasi2016}, \textit{NARM} \cite{Li2017narm} and \textit{STAMP} \cite{Liu2018stamp}. \end{ereview} \\

\hline
\multicolumn{2}{p{8cm}}{Association Rules-based Methods}\\ \hline
\textit{Co-Occurrence (CO)} &  Recommends articles commonly viewed together with the last read article in previous user sessions. This algorithm is a simplified version of the association rules technique, having two as the maximum rule size (pairwise item co-occurrences) (\cite{ludewig2018evaluation,jugovac2018streamingrec}).\\ \hline
\textit{Sequential Rules (SR)} &  The method also uses association rules of size two. It however considers the sequence of the items within a session. A rule is created when an item \textit{q} appeared after an item \textit{p} in a session, even when other items were viewed between \textit{p} and \textit{q}. The rules are weighted by the distance \textit{x} (number of steps) between \textit{p} and \textit{q} in the session with a linear weighting function $ w_{\text{\textsc{SR}}} = 1/x $ \cite{ludewig2018evaluation};\\
\hline
\multicolumn{2}{l}{Neighborhood-based Methods}\\ \hline
\textit{Item-kNN} & Returns the most similar items to the last read article using the cosine similarity between their vectors of co-occurrence with other items within sessions.
This method has been commonly used as a baseline when neural approaches for session-based recommendation were proposed, e.g., in  \cite{hidasi2016}.\\ \hline
\textit{Vector Multiplication Session-Based kNN (V-SkNN)} & This method compares the entire active session with past (neighboring) sessions to determine items to be recommended. The similarity function emphasizes items that appear later within the session. The method proved to be highly competitive in the evaluations in \cite{jannach2017recurrent,ludewig2018evaluation,jugovac2018streamingrec}. \\ \hline
\multicolumn{2}{l}{Other Methods}\\ \hline
\textit{Recently Popular (RP)} & This method recommends the most viewed articles within a defined set of recently observed user interactions on the news portal (e.g., clicks during the last  hour). Such a strategy proved to be very effective in the \textit{2017 CLEF NewsREEL Challenge} \cite{ludmann2017recommending}. \\ \hline
\textit{Content-Based (CB)} &  For each article read by the user, this method suggests recommendable articles with similar content to the last clicked article, based on the cosine similarity of their \textit{Article Content Embeddings}. \\ \hline
\end{tabular}
\begin{tablenotes}
\item[1] \textit{GRU4Rec} v2 \cite{hidasi2018recurrent} was released on Jun 12, 2017 and is available at https://github.com/hidasib/GRU4Rec
\item[2] We exchanged the original negative sampling approaches used for training \textit{GRU4Rec} by the sampling strategy described in Section~\ref{sec:eval_protocol} (i.e., popularity-biased from recent clicks), and observed accuracy improvements for \textit{GRU4Rec} in these experiments.
\end{tablenotes}
\end{threeparttable}
\end{table}

\subsection{Evaluation Methodology}
\label{sec:eval_methodology}
One main goal of our experimental analyses is to make our evaluations as realistic as possible. We therefore did not use the common evaluation approach of random train-test splits and cross-validation. Instead, we use the temporal offline evaluation method that we proposed in \cite{moreira2018news}, which simulates a streaming flow of user interactions (clicks) and new articles being published, whose value quickly decays over time. Since in practical environments it is highly important to very quickly react to incoming events \cite{ludmann2017recommending, kille2017clef}, the baseline recommender methods were constantly updated over time.

\textit{CHAMELEON}'s \textit{NAR} module supports online learning, as it is trained on mini-batches. In our training protocol, we decided to emulate a streaming scenario, in which each user session is used for training only once. Such a scalable approach is different from many model-based recommender systems, like \textit{GRU4Rec} \begin{ereview}and \textit{SR-GNN}\end{ereview}, which require training for some epochs on a large set of recent user interactions to reach competitive accuracy results.

\subsubsection{Evaluation Protocol}
\label{sec:eval_protocol}

\begin{figure}[h!t]
	\centering\includegraphics[width=0.99\linewidth]{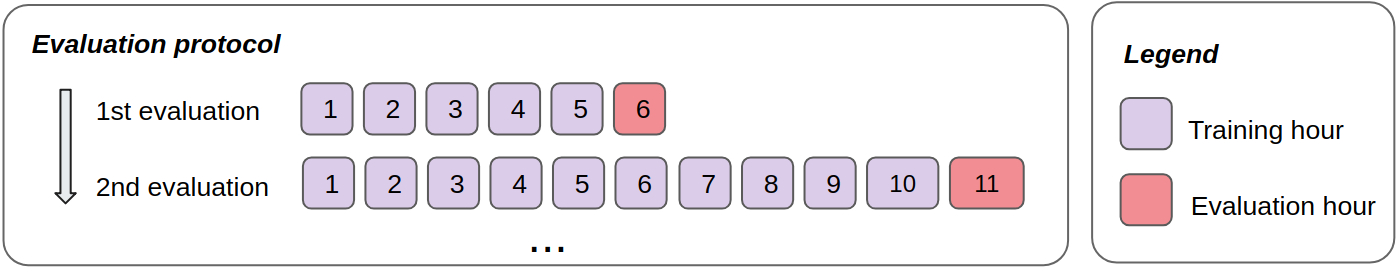}	\caption{Illustration of the evaluation protocol. After training for 5 hours, we evaluate using the sessions of the next hour.}
	\label{figure:eval_protocol}
\end{figure}

The evaluation process works as follows:
\begin{itemize}
\item The recommenders are continuously trained on the users' sessions ordered by time and grouped by hours. Each five hours, the recommenders are evaluated on sessions from the next hour, as exemplified in Fig.~\ref{figure:eval_protocol}. With this interval of five hours (not a divisor of 24 hours), it was possible to sample different hours of the day across the dataset for  evaluation. After the evaluation of the next hour was done, this hour is also  considered for training, until the entire dataset is  covered.\footnote{Our datasets comprises 16 days. We used the first two days to learn an initial model for the session-based algorithms and report the averaged measures after that warm-up period.} It is important to note that, while the most of the baseline methods were continuously updated during the evaluation hour, \begin{ereview}the neural methods---\textit{CHAMELEON}, \textit{SR-GNN}, and \textit{GRU4Rec}---were not trained as evaluation progressed.\footnote{\begin{ereview}Additionally, as the original implementations of \textit{SR-GNN} and \textit{GRU4Rec} do not support fine tuning of previously trained models with \begin{ereview}more data\end{ereview}, those models were trained (for some epochs) considering only sessions from the last 5 hours before each evaluation. On the other hand, \textit{CHAMELEON}'s network was incrementally trained over time (except during evaluation).\end{ereview}} This allows us to emulate a realistic scenario in production where the neural network is trained and deployed once an hour to serve recommendations for the next hour\end{ereview};
\item For each session in the evaluation set, we incrementally ``revealed'' one click after the other to the recommender, as done, e.g., in \cite{hidasi2016} and \cite{quadrana2017personalizing}; 
\item For each click to be predicted, we created a set containing \begin{ereview} 50 randomly sampled recommendable articles \end{ereview} \textit{not} viewed by the user in the session (negative samples), plus the true next article (positive sample), as done in \cite{koren2009} and \cite{Cremonesi2010}. \begin{ereview}The sampling strategy was popularity-biased (i.e., the item sampling probability is proportional to its support), so that strong (popular) negative samples are always present\end{ereview}. We then evaluate the algorithms in the task of ranking those 51 items;
\item Given these rankings, standard information retrieval metrics can be computed.
\end{itemize}

For a realistic evaluation, it is important that the chosen negative samples consist of articles which would be of some interest to readers and which were also available for recommendation in the news portal at a given point of time. For the purpose of this study, we therefore selected as recommendable articles the ones that received at least one click by any user in the preceding hour. To finally select the negative samples, we implemented a popularity-based sampling strategy similar to the one from \cite{hidasi2016}.

\subsubsection{Metrics}
To measure quality factors such as accuracy, item coverage, novelty, and diversity, we have selected a set of top-N metrics from the literature. We chose the cut-off threshold at N=10, representing about 20\% of the list containing the 51 sampled articles (1 positive sample and 50 negative samples). 

The accuracy metrics used in our study were the \textit{Hit Rate (HR@n)}, which checks whether or not the true next item appears in the top-N ranked items, and the \textit{Mean Reciprocal Rank (MRR@n)}, a ranking metric that is sensitive to the position of the true next item in the list. Both metrics are common when evaluating session-based recommendation algorithms \cite{hidasi2016,ludewig2018evaluation,jugovac2018streamingrec}.

As an additional metric, we considered \textit{Item Coverage (COV@n)}, which is sometimes also called ``aggregate diversity'' \cite{adomavicius2011improving}. The idea here is to measure to what extent an algorithm is able to diversify the recommendations and to make a larger fraction of the item catalog visible to the users. 
We compute coverage as the number of distinct articles that appeared in any top-N list divided by the number of recommendable articles \cite{jannach2015recommenders}, i.e., those that were clicked at least once in the last hour.

To measure novelty and diversity, we adapted the evaluation metrics that were proposed in \cite{vargas2011rank, castells2015novelty, vargas2015thesis}. We provide details of their implementation in Appendix~\ref{sec:novelty_diversity_metrics}.
The novelty metrics \textit{ESI-R@n} and \textit{ESI-RR@n} are based on item popularity, returning higher values for long-tail items. The \textit{ESI-R@n} (Expected Self-Information with  Rank-sensitivity) metric includes a rank discount, so that items in the top positions of the recommendation list have a higher effect on the metric. The \textit{ESI-RR@n} (Expected Self-Information with Rank- and Relevance-sensitivity) metric not only considers a rank discount, but also combines novelty with accuracy, as the relevant (clicked) item will have a higher impact on the metric if it is among the top-n recommended items.
Our diversity metrics are based on the \textit{Expected Intra-List Diversity (EILD)} metric. Analogously to the novelty metrics, there are variations to account for rank-sensitivity (\textit{EILD-R@n}) and for both rank- and relevance-sensitivity (\textit{EILD-RR@n}).

For our experiments, all recommender algorithms were tuned towards higher accuracy (\textit{MRR@10}) for each dataset using random search on a hold-out validation set. The resulting best hyper-parameters are reported in Appendix~\ref{sec:hyperparams}.

\section{Results and Discussion}
\label{sec:results}

In this section, we present the main results and discuss our findings under the perspective of our research questions.
For all tables presented in this section, best results for a metric are printed in bold face.
If the best results are significantly different\footnote{As errors around the reported averages were normally distributed, we used paired Student's t-tests with Bonferroni correction for significance tests.}
from measures of all other algorithms, they are marked with *** when $p<0.001$, with ** when $p<0.01$, and with * symbol when $p<0.05$.

\subsection{Evaluation of Recommendation Quality (RQ1)}
\label{sec:results_session_based}
In this section, we first analyze the obtained accuracy results and then discuss the other quality factors.

\subsubsection{Accuracy Analysis}

Table~\ref{tab:accuracy_benchmarks} shows the accuracy results obtained by the different algorithms in terms of the \textit{HR@10} and \textit{MRR@10} metrics. The reported values correspond to the average of the measures obtained for each evaluation hour, according to the evaluation protocol (Section~\ref{sec:eval_methodology}).

\begin{table}[h!t]
\centering
\caption{Accuracy results for \textit{G1} and \textit{Adressa}}
\vspace{10pt}
\footnotesize
\begin{tabular}{lll|ll}

\hline
&
\multicolumn{2}{c}{\textbf{G1 dataset}} &
\multicolumn{2}{c}{\textbf{Adressa dataset}}
\\
\hline
  \textit{Algorithm}
  & \textit{HR@10}
  & \textit{MRR@10}
  & \textit{HR@10}
  & \textit{MRR@10}
 \\

\hline

\textit{CHAMELEON} & \textbf{0.6738}*** & \textbf{0.3458}*** & \textbf{0.7018}*** & \textbf{0.3421}*** \\
\textit{SR} & 0.5900 & 0.2889 & 0.6288 & 0.3022 \\
\textit{Item-kNN} & 0.5707 & 0.2801 & 0.6179 & 0.2819 \\
\textit{CO} & 0.5689 & 0.2626 & 0.6131 & 0.2768 \\
\textit{V-SkNN} & 0.5467 & 0.2494 & 0.6140 & 0.2723 \\
\begin{ereview}\textit{SR-GNN}\end{ereview} & 0.5144 & 0.2467 & 0.6122 & 0.2991 \\
\begin{ereview}\textit{GRU4Rec}\end{ereview} & 0.4669 & 0.2092 & 0.4958 & 0.2200 \\
\textit{RP} & 0.4577 & 0.1993 & 0.5648 & 0.2481 \\
\textit{CB} & 0.3643 & 0.1676 & 0.3307 & 0.1253 \\

\hline
\end{tabular}

\label{tab:accuracy_benchmarks}
\end{table}

In this comparison, our \textit{CHAMELEON} instantiation outperforms the other baseline algorithms on both datasets and on both accuracy metrics by a large margin. The \textit{SR} method performs second-best.

Generally, the observed difference between CHAMELEON and \textit{SR} is higher for the \textit{G1} dataset. This can be explained by the facts that (a) the number of articles in the \textit{G1} dataset is more than 3 times higher than in the other dataset and (b) the \textit{G1} dataset has a lower popularity bias, see the \textit{Gini index} in Table~\ref{tab:datasets}. As a result, algorithms that have a higher tendency to recommend popular items are less effective for datasets with a more balanced click distribution. Looking, for example, at the algorithm that simply recommends recently-popular articles (\textit{RP}), we see that its performance is much higher for the \textit{Adressa} dataset, even though the best obtained measures are almost similar for both datasets.

\begin{ereview}
We can furthermore observe that other neural approaches (i.e., \textit{SR-GNN} and \textit{GRU4Rec}) were not able to provide better accuracy than non-neural baselines for session-based news recommendation.
One of the reasons is that in a real-world scenario---as emulated in our evaluation protocol---those models cannot be updated as often as the baseline methods, due to challenges of asynchronous model training and frequent deployment. 
Furthermore, \textit{CHAMELEON}'s architecture was designed to be able to recommend fresh articles not seen during training. \textit{SR-GNN} and \textit{GRU4Rec} in contrast, cannot make recommendations for items that were not encountered during training, which limits their accuracy in a realistic evaluation. In our datasets, for example, we found that about 3\% (Adressa) to 4\% (G1) of the item clicks in each evaluation hour were on fresh articles, i.e., on articles that were not seen in the preceding training hours.

From the two neural methods, the newer graph-based \textit{SR-GNN} method was performing much better than \textit{GRU4Rec} in our problem setting. 
However, as our detailed analysis in Section~\ref{sec:features_results} will show, \textit{SR-GNN} does not achieve the performance levels of \textit{CHAMELEON}, even when \textit{CHAMELEON} is not leveraging any additional side information other than the article ID (configuration \textit{IC1} in Table \ref{tab:input_types_results}).
\end{ereview}

\begin{figure*}[h!t]
    \centering
    \includegraphics[width=1.0\textwidth]{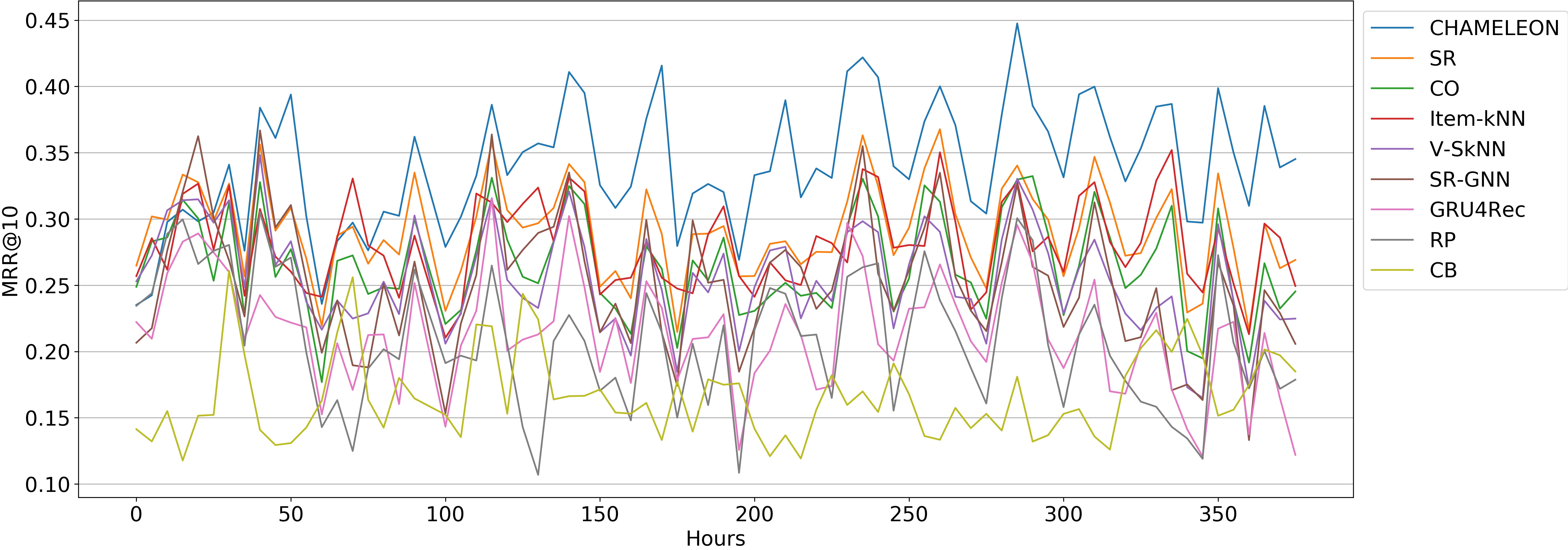}
    \caption{\textit{G1} (16 days) - Detailed results after every 5 hours (\textit{MRR@10})}
    \label{fig:accuracy_over_time_g1}
\end{figure*}

\begin{figure*}[h!t]
    \centering
    \includegraphics[width=1.0\textwidth]{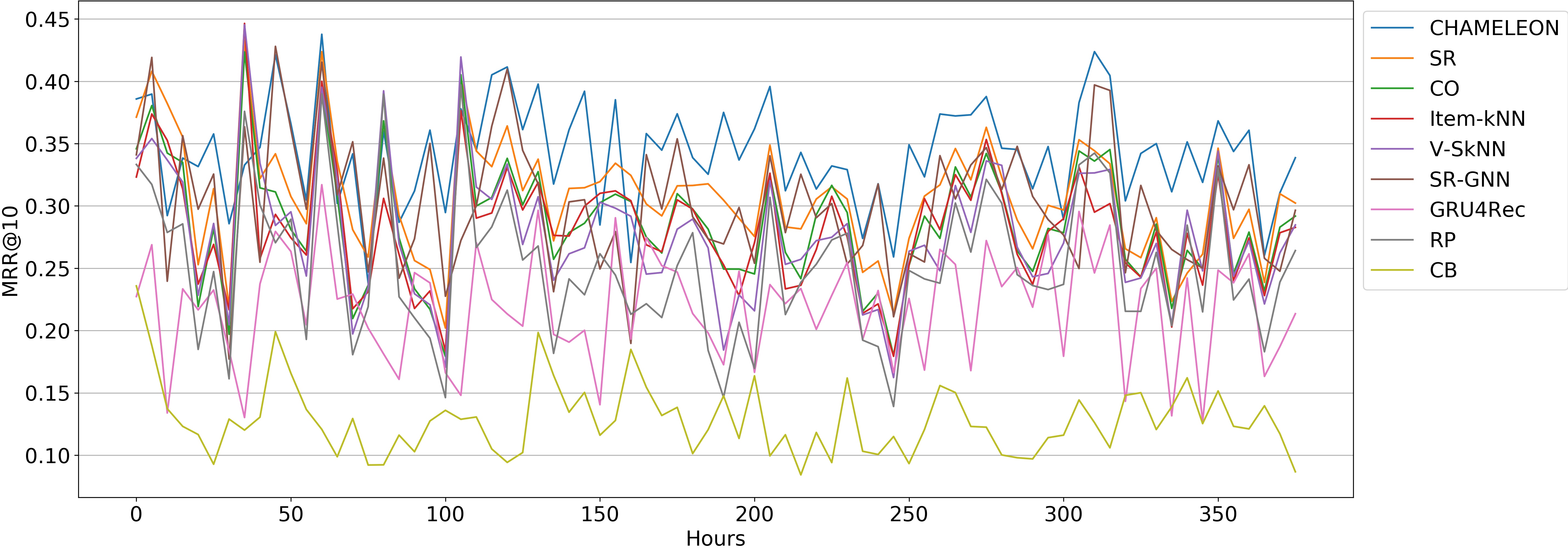}
    \caption{\textit{Adressa} (16 days) - Detailed results after every 5 hours (\textit{MRR@10})}
    \label{fig:accuracy_over_time_adressa}
\end{figure*}

In Fig.~\ref{fig:accuracy_over_time_g1}~and~\ref{fig:accuracy_over_time_adressa}, we plot the obtained accuracy values (\textit{MRR@10}) of the different algorithms along the 16 days, with an evaluation after every 5 hours. We can note that, after some training hours, \textit{CHAMELEON} clearly recommends with higher accuracy than all other algorithms.

\subsubsection{Analysis of Additional Quality Factors}
The results obtained for the other recommendation quality factors investigated in our research---item coverage, novelty, and diversity---are shown in Table~\ref{tab:other_factors_benchmarks}. The observations can be summarized as follows:

\begin{table}[h!t]
\centering
\caption{Evaluation of other quality factors for the \textit{G1} and \textit{Adressa} datasets}
\label{tab:other_factors_benchmarks}
\footnotesize
\begin{tabular}{p{1.4cm}p{1.1cm}p{0.9cm}p{0.9cm}p{0.9cm}p{0.9cm}}
\hline
 & \textit{Item Coverage}
 & \multicolumn{2}{c}{\textit{Novelty}}
 & \multicolumn{2}{c}{\textit{Diversity}} \\
 \textit{Recommender}
 & \textit{COV@10}
 & \textit{ESI-R@10}
 & \textit{ESI-RR@10}
 & \textit{EILD-R@10}
 & \textit{EILD-RR@10}
 \\
\hline
\multicolumn{6}{p{8cm}}{\textbf{G1 dataset}}\\ \hline
\textit{CHAMELEON} & 0.6373 & 6.4177 & \textbf{0.7302}*** & 0.3620 & \textbf{0.0419}*** \\
\textit{SR} & 0.2763 & 5.9747 & 0.5747 & 0.3526 & 0.0374 \\
\textit{Item-kNN} & 0.3913 & 6.5909 & 0.6301 & 0.3552 & 0.0361 \\
\textit{CO} & 0.2499 & 5.5728 & 0.5126 & 0.3570 & 0.0352 \\
\textit{V-SkNN} & 0.1355 & 5.1760 & 0.4411 & 0.3558 & 0.0339 \\
\begin{ereview}\textit{SR-GNN}\end{ereview} & 0.3196 & 5.4280 & 0.5093 & 0.3668 & 0.0350  \\
\begin{ereview}\textit{GRU4Rec}\end{ereview} & 0.6333 & 5.2332 & 0.3925 & 0.3662 & 0.0310  \\
\textit{RP} & 0.0218 & 4.4904 & 0.3259 & \textbf{0.3750}*** & 0.0296 \\
\textit{CB} & \textbf{0.6774} & \textbf{8.1531}*** & 0.5488 & 0.2789 & 0.0193 \\
\hline
\multicolumn{6}{p{8cm}}{\textbf{Adressa dataset}}\\ \hline
\textit{CHAMELEON} & 0.7926 & 5.3410 & \textbf{0.6083}*** & 0.2123 & \textbf{0.0250}*** \\
\textit{SR} & 0.4604 & 5.4443 & 0.5277 & 0.2188 & 0.0235 \\
\textit{CO} & 0.4220 & 5.0789 & 0.4748 & 0.2138 & 0.0222 \\
\textit{Item-kNN} & 0.5314 & 5.4675 & 0.5091 & \textbf{0.2246} & 0.0228 \\
\textit{V-SkNN} & 0.1997 & 4.6018 & 0.4112 & 0.2112 & 0.0217 \\
\begin{ereview}\textit{SR-GNN}\end{ereview} & 0.5197 & 5.1013 & 0.5125 & 0.2214 & 0.0241 \\
\begin{ereview}\textit{GRU4Rec}\end{ereview} & 0.5143 & 5.0571 & 0.3782 & 0.2131 & 0.0184  \\
\textit{RP} & 0.0542 & 4.1465 & 0.3486 & 0.2139 & 0.0200 \\

\textit{CB} & \textbf{0.8875}*** & \textbf{7.6715}*** & 0.4104 & 0.0960 & 0.0060 \\

\hline

\end{tabular}

\end{table}

\begin{itemize}
\item  In terms of \textit{item coverage (COV)}, \textit{CHAMELEON} has a much richer spectrum of articles that are included in its top-10 recommendations compared to other algorithms, suggesting a higher level of personalization. The only method with a higher coverage was the \textit{CB} method, which however is not very accurate. This is expected for a method that is agnostic of an article's popularity.
\item Looking at \textit{novelty}, the \textit{CB} method also recommends the least popular, and thus more novel articles, according to the \textit{ESI-R} metric. This effect has been observed in other works such as \cite{castells2015novelty, celma2008new}, which is expected as this is the only method that does not take item popularity into account in any form. \textit{CHAMELEON} ranks third on this metric for the \textit{G1} dataset and is comparable to the other algorithms for \textit{Adressa}\footnote{We will show later, in Section~\ref{sec:tradeoff_results}, how the novelty of \textit{CHAMELEON} can be increased based on the novelty regularization method proposed in Section~\ref{sec:new_loss_function}.}. Looking at novelty in isolation is, however, not sufficient, which is why we include the relevance-weighted \textit{ESI-RR} metric as well. When novelty and relevance are combined in one metric, it turns out that CHAMELEON leads to the best values on both datasets-
\item Considering \textit{diversity}, we can observe that most algorithms are quite similar in terms of the \textit{EILD-R@10} metric. The \textit{CB} method has the lowest diversity by design, as it always recommends articles with similar content. When article relevance is taken into account along with diversity with the \textit{EILD-RR@10} metric, we again see that \textit{CHAMELEON} is more successful than others in balancing diversity and accuracy.
\end{itemize}

\subsection{Analyzing the Importance of Input Features for the NAR module (RQ2)}

\label{sec:features_results}
\textit{CHAMELEON} leverages a number of input features to provide more accurate recommendations, as shown in Table~\ref{tab:nar-features}. In order to understand the effects of including those features in our model, we performed a number of additional experiments with features combined in different Input Configurations (IC)\footnote{This process is sometimes referred to as \textit{ablation study}.}. Table~\ref{tab:input_config} shows five different configurations where we start only with the article IDs (\textit{IC1}) and incrementally add more features until we have the  model with all input features (\textit{IC5}).

Note that we have included two variations of \textit{IC3}: (a) using the \textit{Article Content Embedddings (ACE)} learned with the \textit{ACR} module, trained to predict article metadata attributes from text (supervised learning), and (b) using article embeddings trained with \textit{doc2vec} \cite{le2014distributed} (unsupervised learning).

\begin{table}[!htbp]
\centering
\caption{\textit{Input Configurations (IC)} for the \textit{NAR} module}
\label{tab:input_config}
\vspace{10pt}
\footnotesize
\begin{tabular}{lp{6cm}}
\hline
 \textbf{Input config.}
 & \textbf{Feature Sets}
 \\ \hline
\textit{IC1} & Article Id \\
\textit{IC2} & \textit{IC1} + Article Context (Novelty and Recency) \\
\textit{IC3} (ACE) & \textit{IC2} + Article Content represented as the \textit{Article Content Embeddings} learned by the \textit{ACR module} \\
\textit{IC3} (doc2vec) & \textit{IC2} + Article Content represented as \textit{doc2vec} embeddings \\
\textit{IC4} & \textit{IC3} + Article Metadata \\
\textit{IC5} & \textit{IC4} + User Context \\
\hline
\end{tabular}
\end{table}

Table~\ref{tab:input_types_results} shows the results of this study. 
We can generally see that both accuracy (\textit{HR@10} and \textit{MRR@10}) and item coverage (\textit{COV@10}) improve when more input features are considered in the \textit{NAR} module.
The largest improvements in terms of accuracy for both datasets can be observed when the feature set \textit{Article Metadata} (\textit{IC3}) are included. The feature sets of \textit{User Context} (\textit{IC5}) and \textit{Article Context} (\textit{IC2}) also played an important role when generating the recommendations.

We can also observe cases where measures become lower with the addition of new features. For both datasets, for example, the diversity of \textit{CHAMELEON}'s recommendations in terms of the \textit{EILD-R} metric decreases with additional features, in particular when the \textit{Article Content} features is included at \textit{IC3}. This is expected, as recommendations become generally more similar when content features are used in a hybrid RS.

Looking at the two variations of configuration \textit{IC3}, we can observe that for the \textit{G1} dataset the textual content representation of \textit{ACE} leads to a much higher accuracy than \textit{doc2vec} embeddings. This confirms the usefulness of our specific way of encoding the textual content with the \textit{ACR} module, based on word embeddings pre-trained in a larger corpus (e.g. Wikipedia).

For the \textit{Adressa} dataset, however, the results with \textit{ACE} and \textit{doc2vec} are very similar\footnote{Except for the \textit{EILD-R@10} metric, which cannot be compared because this metric uses different content embeddings (\textit{ACE} or \textit{doc2vec}) to compute similarities in this case.}. A possible explanation for the difference between the datasets can lie in the nature of the available metadata of the articles, which are used as target attributes during training. In the \textit{G1} dataset, for example, we have 461 article categories, which is much more than for the \textit{Adressa} dataset, with 41 categories. Furthermore, the distribution of articles by category is more unbalanced for \textit{Adressa} (Gini index = 0.883) than for \textit{G1} (Gini index = 0.820). In theory, fine-grained metadata can lead to content embeddings clustered around distinctive topics, which may be useful to recommend related content.

\begin{table}[h!t]
\centering
\caption{Effects of different input feature configurations on recommendation quality.}
\label{tab:input_types_results}
\footnotesize
\begin{tabular}{lp{0.9cm}p{0.9cm}p{0.9cm}p{0.9cm}p{0.9cm}}
\hline

 \textit{Recommender}
 & \textit{HR@10}
 & \textit{MRR@10}
 & \textit{COV@10}
 & \textit{ESI-R@10}
 & \textit{EILD-R@10} \\
\hline  \hline
\multicolumn{6}{p{8cm}}{\textbf{G1 dataset}}\\ \hline

\textit{IC1} & 0.5708 & 0.2674 & 0.6084 & 6.2597 & \textbf{0.4515} \\
\textit{IC2} & 0.6073 & 0.2941 & 0.6095 & 6.1841 & 0.3736 \\
\hline
\textit{IC3 (doc2vec)} & 0.6169 & 0.3003 & 0.6211 & 6.2115 & (0.4504) \\
\textit{IC3 (ACE)} & 0.6472 & 0.3366 & 0.6296 & 6.1507 & 0.3625 \\
\hline
\textit{IC4} & 0.6483 & 0.3397 & 0.6316 & 6.1573 & 0.3621 \\
\textit{IC5} & \textbf{0.6738}*** & \textbf{0.3458}* & \textbf{0.6373} & \textbf{6.4177}** & 0.3620 \\

\hline  \hline
\multicolumn{6}{p{8cm}}{\textbf{Adressa dataset}}\\ \hline

\textit{IC1} & 0.6779 & 0.3260 & 0.7716 & 5.3296 & \textbf{0.2190} \\
\textit{IC2} & 0.6799 & 0.3273 & \textbf{0.8034} & 5.2636 & 0.2187 \\
 \hline
\textit{IC3 (doc2vec)} & 0.6907 & 0.3339 & 0.7951 & 5.2856 & (0.4565) \\
\textit{IC3 (ACE)} & 0.6906 & 0.3348 & 0.7820 & 5.2771 & 0.2103 \\
 \hline
\textit{IC4} & 0.6906 & 0.3362 & 0.7882 & 5.2900 & 0.2123 \\
\textit{IC5} & \textbf{0.7018}*** & \textbf{0.3421}**  & 0.7926 & \textbf{5.3410} & 0.2123 \\
\hline
\end{tabular}

\end{table}

\subsection{Balancing Accuracy and Novelty with CHAMELEON (RQ3)}
\label{sec:tradeoff_results}
In this section, we analyze the effectiveness of our novel technical approach to balance accuracy and novelty within \textit{CHAMELEON}, as described in Section \ref{sec:new_loss_function}. Specifically, we conducted a sensitivity analysis for the novelty regularization factor ($\beta$) in the proposed loss function.

\begin{table}[h!t]
\centering
\caption{Evaluation of \textit{CHAMELEON}'s loss regularization factor for novelty ($ \beta $)}
\label{tab:nov_div_sensitivity}
\vspace{10pt}
\footnotesize

\begin{tabular}{rlll}
\hline
 \textit{Reg. factors}
 & \textit{ESI-R@10}
 & \textit{MRR@10}
 & \textit{COV@10}
 \\
\hline
\multicolumn{4}{p{6cm}}{\textbf{G1 dataset}}\\ \hline
$ \beta=0.0 $  & 6.4177 & \textbf{0.3458} & 0.6373 \\
$ \beta=0.1 $  & 6.9499 & 0.3401 & 0.6785 \\
$ \beta=0.2 $  & 7.7012 & 0.3222 & 0.6962 \\
$ \beta=0.3 $  & 8.5763 & 0.2933 & 0.7083 \\
$ \beta=0.4 $  & 9.3054 & 0.2507 & 0.7105 \\
$ \beta=0.5 $  & \textbf{9.8012} & 0.2170 & \textbf{0.7123}* \\

\hline
\multicolumn{4}{p{6cm}}{\textbf{Adressa dataset}}\\ \hline
$ \beta=0.0 $  & 5.3410 & \textbf{0.3421} & 0.7926 \\
$ \beta=0.1 $  & 5.8279 & 0.3350 & 0.8635 \\
$ \beta=0.2 $  & 7.5561 & 0.2948 & 0.9237 \\
$ \beta=0.3 $  & 9.4709 & 0.2082 & 0.9353 \\
$ \beta=0.4 $  & 10.2500 & 0.1560 & \textbf{0.9376} \\
$ \beta=0.5 $  & \textbf{10.5184} & 0.1348 & 0.9365 \\
\hline
\end{tabular}

\end{table}

\begin{figure}[h!t]
  \begin{subfigure}[b]{0.5\textwidth}
    \includegraphics[width=0.90\textwidth]{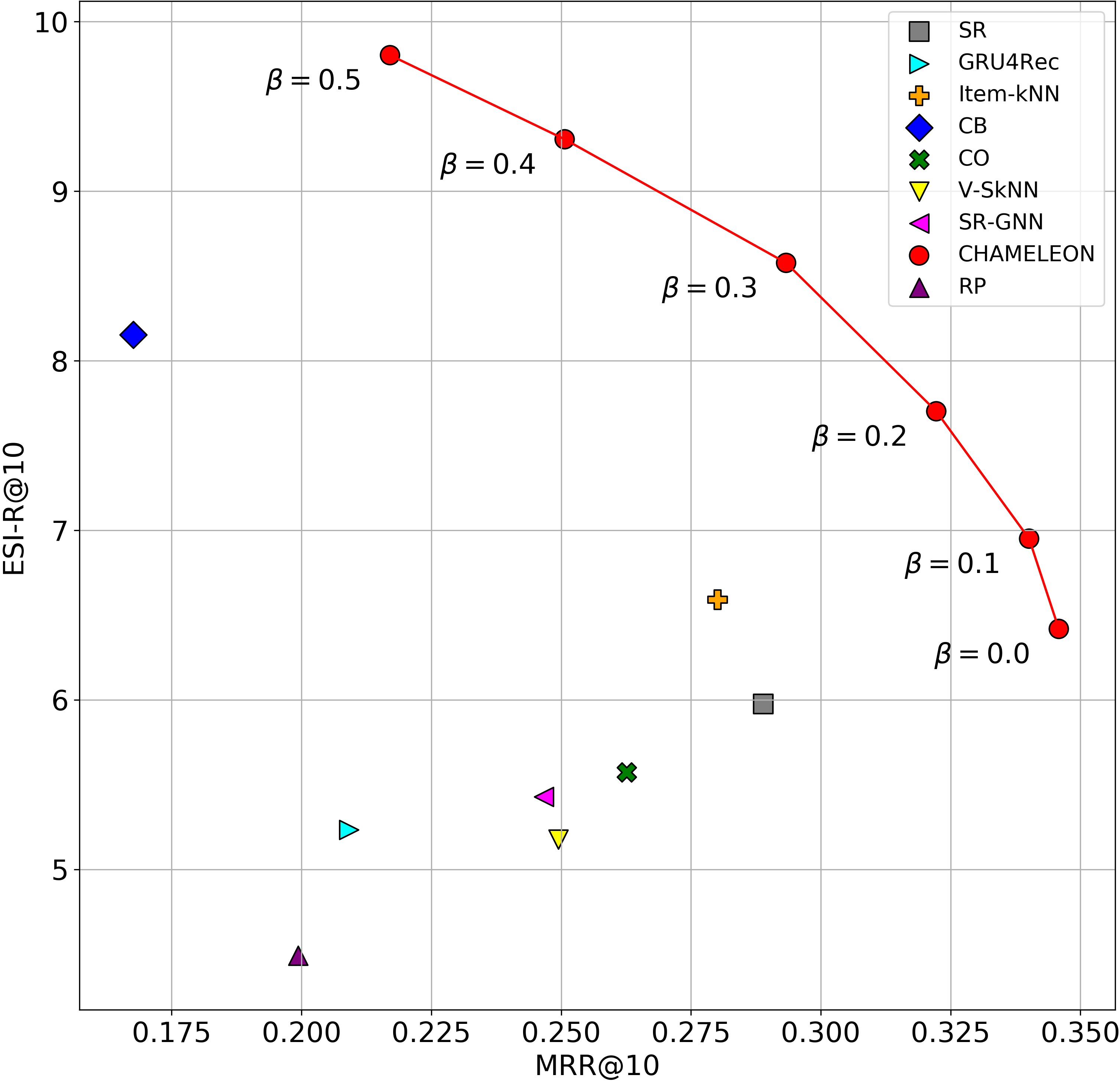}
    \caption{\begin{ereview}\textit{G1} dataset\end{ereview}}
    \label{fig:tradeoff_novelty_g1}
  \end{subfigure}
  \begin{subfigure}[b]{0.5\textwidth}
    \includegraphics[width=0.90\textwidth]{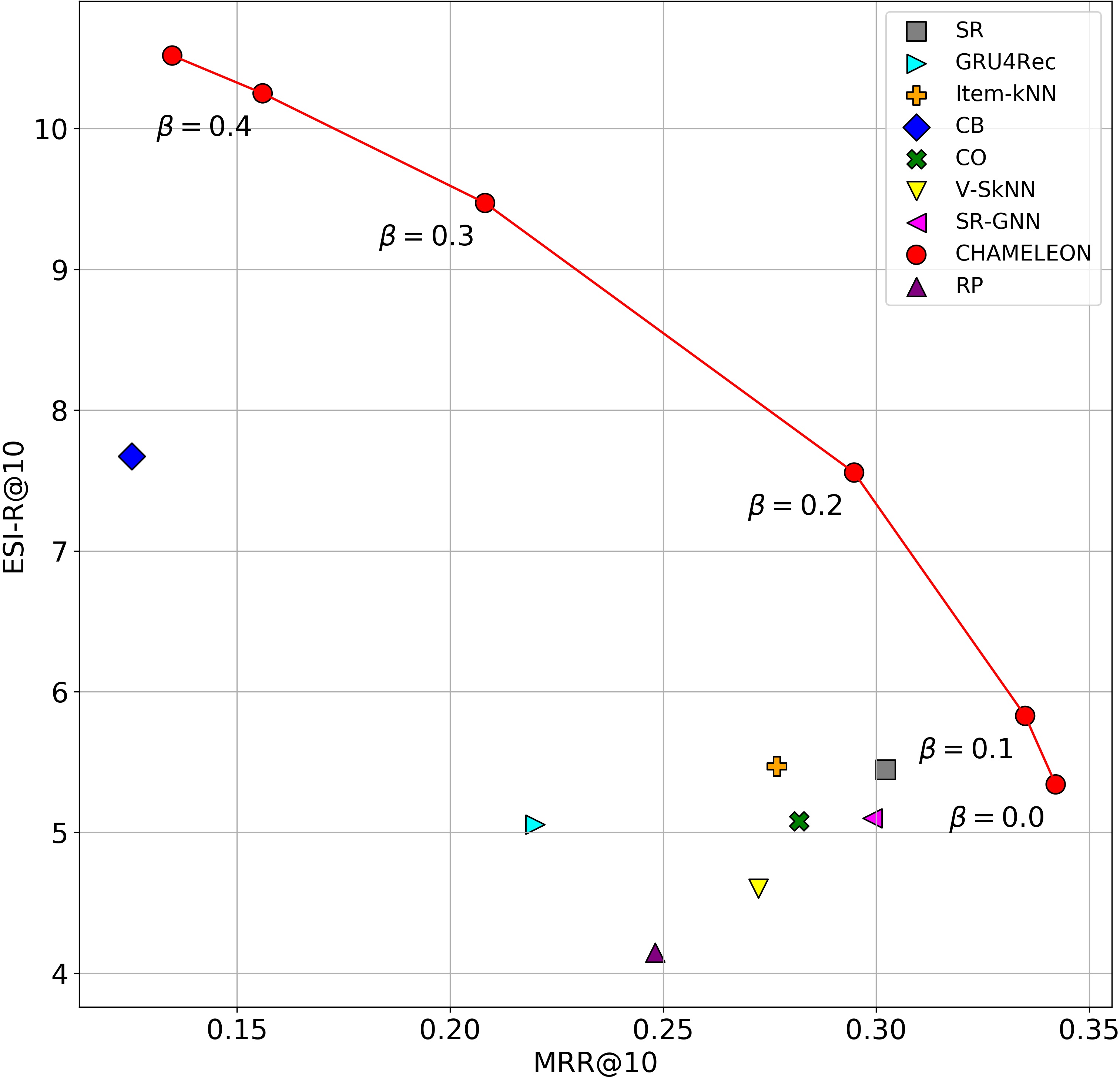}
    \caption{\begin{ereview}\textit{Adressa} dataset\end{ereview}}
    \label{fig:tradeoff_novelty_adressa}
  \end{subfigure}
  \caption{Trade-off between Accuracy (MRR@10) and Novelty (ESI-R) for different values of $\beta$.}
  \label{fig:tradeoff_novelty}
\end{figure}

Table~\ref{tab:nov_div_sensitivity} shows the detailed outcomes of this analysis. As expected, increasing the value of $\beta$  increases the novelty of the recommendations and also leads to higher item coverage. Correspondingly, the accuracy values decrease with higher levels of novelty.  Fig.~\ref{fig:tradeoff_novelty} shows a scatter plot that illustrates some effects and contrasts of the obtained results in our evaluation. The trade-off between accuracy (\textit{MRR@10}) and novelty (\textit{ESI-R@10}) for \textit{CHAMELEON} can be clearly identified. We also plot the results for the baseline methods here for reference. This comparison reveals that tuning  $\beta$ helps us to end up with recommendations that are both more accurate and more novel than the ones by the baselines. 
Fig.~\ref{fig:tradeoff_novelty} also illustrates the differences between the two datasets. Due to the uneven distribution of the \textit{Adressa} dataset, the performance improvements over the \textit{RP} baseline, which recommends recently popular items, are smaller than for the \textit{G1} dataset.

\section{Summary and Future Works}
\label{sec:summary-and-outlook}
\begin{ereview}
In this final section, we first summarize the \begin{ereview}major\end{ereview} findings of our work and then give an outlook on future research directions in this area.
\end{ereview}

\subsection{Summary}
\begin{ereview}
We have proposed a novel approach for session-based news recommendation, which in particular addresses domain-specific problems such as a) the short lifetime of the recommendable items and b) the lack of longer-term preference profiles of the users. The main technical contribution of our work lies in the combination of \emph{content} and \emph{context} features and a sequence modeling technique based on Recurrent Neural Networks. Furthermore, we propose a novel way to balance potentially conflicting optimization goals like accuracy and novelty through a parameterizable loss function.

The individual technical components that were developed in our work were integrated into a configurable open-source news recommendation framework for session-based recommendations. Experimental evaluations on two public news datasets revealed that a) the proposed hybrid approach leads to higher prediction accuracy and b) that our approach to balance conflicting optimization goals is effective.

\end{ereview}





\subsection{Future Works}
With respect to future works, our plan is to further investigate differences between existing algorithms in terms of their capability of dealing with the constant item cold-start problem, which is omnipresent in news portals.

\begin{ereview}Another specific challenge that we have not addressed so far and which was not investigated to a large extent in the literature as well is that of ``outliers'' in the user profiles. Specifically, there might be a certain level of noise in the user profiles. In the case of news recommendation, this could be random clicks by the user or user actions that result from a click-bait rather than from genuine user interest. As proposed in previous works \cite{Saiaetal2016asemantic,Said:2018:CIR:3231208.3231212,DBLP:conf/ic3k/SaiaBC14}, we plan to identify such outliers and noise in the context of session-based recommendation to end up with a better estimate of the true user intent within a session.\end{ereview}

\begin{ereview}
Furthermore, we will investigate the role of emotions as a further contextual factor, see, e.g. \cite{POIRSON2019281,Mizgajski2019}, both in the form of trying to consider the sentiment of a given news article and the current emotional state of the user.
\end{ereview}

Finally, our next immediate goals include the exploration of mechanisms within \emph{CHAMELEON} that allow us to balance more than two quality factors, with a particular look at enhancing the diversity of the recommendations while preserving accuracy.

\section*{Acknowledgments}
G. Moreira thanks CI\&T for supporting this research in its R\&D departments (D1 / Lab23), Globo.com for sharing a dataset and their technical challenges, and also Ecossistema Negocios Digitais Ltda for their support for this article.

\appendices

\section{Novelty and Diversity metrics}
\label{sec:novelty_diversity_metrics}
In our studies, we use novelty and diversity metrics adapted from \cite{vargas2011rank} and \cite{vargas2015thesis}, which we tailored to fit our specific problem of session-based news recommendation. Generally, for the purpose of this investigation, novelty is evaluated in terms of \textit{Long-Tail Novelty}. Items with high novelty correspond to long-tail items, i.e., items that were clicked on by few users, whilst low novelty items correspond to more popular items.

\subsection{ESI-R@n}

The \textit{Expected Self-Information with Rank-sensitivity} metric, presented in \eqref{eq:esi-r}, was adapted from the \textit{MSI} metric proposed by \cite{castells2015novelty} with the addition of a rank discount. The term \begin{ereview}  $ -\text{log}_2{p(i)} $ \end{ereview} represents the core of this metric, which comes from the \textit{self-information} (also known as \textit{surprisal}) metric of Information Theory, which quantifies the amount of information conveyed by the observation of an event \cite{castells2015novelty}. Applying the \begin{ereview} $ \text{log}(\cdot) $ \end{ereview} function emphasizes the effect of highly novel items. We define $ L = {i_1, ... , i_N} $ as a recommendation list of size $ N = | L | $.

\begin{ereview}
\begin{equation} \label{eq:esi-r}
\text{ESI-R}(L) = \frac{1}{\sum_{j=1}^N \text{disc}(j)}
\sum_{k=1}^N  -\text{log}_2{p(i_k)} \times \text{disc}(k)
\end{equation}
\end{ereview}

In this setting, the probability $ p(i) $ of an item being part of a random user interaction under free discovery is the normalized recent popularity, i.e., $ p(i) = \text{rec\_norm\_pop}(i) $, previously presented in \eqref{eq:rec-norm-pop}.
In \eqref{eq:esi-r}, $ \text{disc}(\cdot) $ is a logarithmic rank discount, defined in \eqref{eq:disc}, that maximizes the impact of novelty for top ranked items, under the assumption that their characteristics will be more visible to users compared to the rest of the top-n recommendation list:

\begin{ereview}
\begin{equation} \label{eq:disc}
\text{disc}(k) = \frac{1}{\text{log}_2(k+1)}
\end{equation}
\end{ereview}

\subsection{ESI-RR@n}
Analyzing quality factors like accuracy, novelty, and diversity in isolation can be misleading. Some Information Retrieval (IR) metrics, such as $ \alpha\!\!-\!\!\text{nDCG} $, therefore  consider novelty contributions only for relevant items for a given query \cite{castells2015novelty}.
As proposed by \cite{vargas2011rank}, a relevance-sensitive novelty metric should likewise assess the novelty level based on the recommended items that are actually relevant to the user.

Thus, we used a variation of a novelty metric to account for relevance---\textit{Expected Self-Information with Rank- and Relevance-sensitivity (ESI-RR@n)}. It weights the novelty contribution by the relevance of an item for a user $ p(rel|i,u)$ \cite{castells2015novelty}. We adapt the proposal from \cite{vargas2015thesis}:

\begin{equation} \label{eq:p_rel}
p(rel|i,u) = \text{relevance}(i,u) =
\begin{cases}
      1.0, & \text{if}\ i \in I_u \\
      b, & \text{otherwise}
    \end{cases},
\end{equation}

where $ I_u $ is the set of items the user interacted within the ongoing session, and $ b $ is a background probability of an unobserved interaction (negative sample) being also somewhat relevant for a user. The lower the value of $ b $ (e.g., $ b=0 $) the higher the influence of relevant items (accuracy) in this metric. The author of \cite{vargas2015thesis} used an empirically determined value of $ b = 0.2 $, based on his experiments on balancing diversity and novelty. In our study, we arbitrarily set $ b = 0.02 $, so that all the 50 negative samples would sum up to the same relevance (1.0) of a positive (clicked) item.

Equation~\eqref{eq:esi-rr} shows how we compute the \textit{ESI-RR@n} metric. 
\begin{ereview}
\begin{equation} \label{eq:esi-rr}
\text{ESI-RR}(L) =
 C_k \sum_{k=1}^N  -\text{log}_2{p(i_k)} \times \text{disc}(k) \times \text{relevance}(i_k,u),
\end{equation}
\end{ereview}

Equation~\eqref{eq:rr-normalization-term} defines the term $ C_k $, which computes the weighted average based on ranking discount.

\begin{equation} \label{eq:rr-normalization-term}
C_k = \frac{1}{\sum_{k' = 1}^N \text{disc}(k')}
\end{equation}

Like in \cite{vargas2015thesis}, the relevance is not normalized, so that more relevant items among the top-n recommendations lead to a global higher novelty.

\subsection{EILD-R@n}
Diversity was measured based on the \textit{Expected Intra-List Diversity} metric proposed by \cite{vargas2011rank}, with variations to account for rank-sensitivity (\textit{EILD-R@n}) and for both rank- and relevance-sensitivity (\textit{EILD-RR@n}).

Intra-List Diversity measures the dissimilarity of the recommended items with respect to the other items in the recommended list. In our case, the distance metric $ d(\cdot) $ defined in \eqref{eq:cos_dist} is the cosine distance.
\begin{equation} \label{eq:cos_dist}
d(a,b) = (1 - \text{sim}(a,b))  / 2,
\end{equation}
Here, $a$ and $b$ are the \textit{Article Content Embeddings} of two articles and $ \text{sim}(a,b) $ is their cosine similarity. As the cosine similarity ranges from -1 to +1, the cosine distance is scaled to the range [0,1].

The \textit{Expected Intra-List Diversity with Rank-sensitivity (EILD-R@n)} metric, defined in \eqref{eq:eild-r}, is the average intra-distance between items pairs weighted by a logarithmic rank discount $ \text{disc}(\cdot) $, defined in \eqref{eq:disc}. Given a recommendation list $ L = {i_1, ... , i_N} $ of size $ N = | L | $, we compute the \textit{EILD-R@n} metric as follows.

\begin{ereview}
\begin{equation} \label{eq:eild-r}
\begin{split}
\text{EILD-R}(L) = &
\frac{1}{\sum_{k'=1}^N \text{disc}(k')} \\&
\sum_{k=1}^N \text{disc}(k) \frac{1}{ \sum_{l'=1:l' \neq k}^N \text{rdisc}(l', k) } \\&
\sum_{l=1:l \neq k}^N d(i_k, i_l) \times  \text{rdisc}(l, k)
\end{split}
\end{equation}
\end{ereview}

The term $ \text{rdisc}(l, k) $, defined in \eqref{eq:rdisc}, represents a relative ranking discount, considering that an item $ l $ that is ranked before the target item $ k $ has already been discovered. In this case, items ranked after $ k $ are assumed to lead to a decreased diversity perception as the relative rank between $ k $ and $ l $ increases.

\begin{equation} \label{eq:rdisc}
\text{rdisc}(l, k) = \text{disc}(\text{max}(0, l-k))
\end{equation}

\subsection{EILD-RR@n}
The \textit{Expected Intra-List Diversity with Rank- and Relevance-sensitivity} finally measures
the average diversity between item pairs, weighting items by rank discount and relevance, analogously to the \textit{ESI-RR@n} metric:

\begin{equation} \label{eq:eild-rr}
\begin{split}
\text{EILD-RR}(L) = &
C_k \sum_{k=1}^N \text{disc}(k) \times \text{relevance}(i_k,u) C_l \\&
\sum_{l=1:l \neq k}^N d(i_k, i_l) \text{rdisc}(k, l) \times  \text{relevance}(i_l,u)
\end{split}
\end{equation}

Here, $ C_k $ \eqref{eq:rr-normalization-term} and $ C_l $ \eqref{eq:eild-internal-normalization-term} are normalization terms representing a weighted average based on rank discounts.

\begin{equation} \label{eq:eild-internal-normalization-term}
C_l = \frac{1}{ \sum_{l'=1:l \neq k}^N \text{rdisc}(k, l') }
\end{equation}

\section{Final Algorithms Hyper-Parameters}
\label{sec:hyperparams}
In Table~\ref{tab:hyperparams}, we present the best hyper-parameters found for each algorithm and dataset. They were tuned for accuracy (\textit{MRR@10}) on a hold-out validation set, by running random search within defined ranges for each hyper-parameter. The methods \textit{CO}, \textit{RP}, and \textit{CB} do not have hyper-parameters. More information about the hyper-parameters can be found in the shared code and in the papers where the baseline methods were proposed.

\begin{table*}[!htbp]
\centering
\caption{Best hyper-parameters per algorithm and dataset}
\label{tab:hyperparams}
\vspace{10pt}
\footnotesize
\begin{tabular}{lp{4.0cm}p{8.0cm}rr}
\hline
 \textit{Method} &
 \textit{Parameter} &
 \textit{Description} &
 \textit{G1} &
 \textit{Adressa}
 \\
\hline \hline

\textit{CHAMELEON} & batch\_size & Number of sessions considered for each mini-batch & 256 & 64 \\
 & learning\_rate & Learning rate for each training step (mini-batch) & 1e-4 & 3e-4 \\
 & reg\_l2 & \textit{L2} regularization of the network's parameters & 1e-5 & 1e-4 \\
 & softmax\_temperature & Used to control the ``temperature'' of the softmax function & 0.1 & 0.2 \\
 & CAR\_embedding\_size & Size of the User-Personalized Contextual Article Embedding & 1024 & 1024 \\
 & rnn\_units & Number of units in an RNN layer & 255 & 255 \\
 & rnn\_num\_layers & Number of stacked RNN layers & 2 & 2 \\

\hline

\textit{SR} & max\_clicks\_dist & Maximum number of clicks to walk back in the session from the currently viewed item. & 10 & 10 \\
 & dist\_between\_clicks\_decay & Decay function for the distance between two items clicks within a session (e.g., linear, same, div, log, quadratic) & div & div \\
\hline

\textit{Item-kNN} & reg\_lambda & Smoothing factor for the popularity distribution to normalize item vectors for co-occurrence similarity & 20 & 20 \\
 & alpha & Balance between normalizing with the support counts of the two items. 0.5 gives cosine similarity, 1.0 gives confidence. & 0.75 & 0.50 \\
\hline

\textit{V-SkNN} & sessions\_buffer\_size & Buffer size of last processed sessions & 3000 & 3000 \\
 & candidate\_sessions\_sample\_size & Number of candidates near the sessions to sample & 1000 & 2000 \\
  & nearest\_neighbor\_session\_for\_scoring & Nearest neighbors to compute item scores & 500 & 500 \\
  & similarity & Similarity function (e.g., Jaccard, cosine) & cosine & cosine \\
  & sampling\_strategy & Strategy for sampling (e.g., recent, random) & recent & recent \\
  & first\_session\_clicks\_decay & Decays the weight of first user clicks in active session when finding neighbor sessions (e.g. same, div, linear, log, quadratic) & div & div \\
\hline

\begin{ereview}\textit{SR-GNN}\end{ereview} & batch\_size & Batch size & 128 & 128 \\
 & n\_epochs & Number of training epochs & 10 & 10 \\
 & hidden\_size & Number of units on hidden state & 200 & 200 \\
 & l2\_lambda & Coefficient of the $ L_2 $ regularization & 1e-5 & 2e-5 \\
 & propagation\_steps & GNN propagation steps  & 1 & 1 \\
 & learning\_rate & Learning rate & 1e-3 & 1e-3 \\
 & learning\_rate\_decay & Learning rate decay factor & 0.15 & 0.1 \\
 & learning\_rate\_decay\_steps & number of steps after which the learning rate decays & 3 & 3 \\
 & nonhybrid & Enables/disables the Hybrid mode & True & True \\

\hline
\begin{ereview}\textit{GRU4Rec}\end{ereview} & batch\_size & Batch size & 128 & 128 \\
 & n\_epochs & Number of training epochs & 3 & 3 \\
 & optimizer & Training optimizer & Adam & Adam \\
 & loss & The loss type & bpr-max-0.5 & bpr-max-0.5 \\
 & layers & Number of GRU units in the layers & [300] & [300] \\
 & dropout\_p\_hidden & Dropout rate & 0.0 & 0.0 \\
 & learning\_rate & Learning rate & 1e-4 & 1e-4 \\
 & l2\_lambda & Coefficient of the $ L_2 $ regularization  & 1e-5 & 2e-5 \\
 & momentum & if not zero, Nesterov momentum will be applied during training with the given strength & 0 & 0 \\
 & embedding & Size of the embedding used, 0 means not to use embedding & 0 & 0 \\

\hline
\hline
\end{tabular}

\end{table*}

\newpage

\bibliography{IEEEabrv,references}

\EOD

\end{document}